\documentclass[twocolumn,aps,superscriptaddress,showpacs,floatfix]{revtex4}
\usepackage{amssymb}
\usepackage{amsmath}
\usepackage{graphicx}
\usepackage[normalem]{ulem}
\usepackage[dvips]{color}
\usepackage{bm}
\usepackage{longtable}

\begin{document}

\title{Single-nucleon potential decomposition of the nuclear symmetry energy}
\author{Rong Chen}
\affiliation{Department of Physics, Shanghai Jiao Tong University, Shanghai 200240, China}
\author{Bao-Jun Cai}
\affiliation{Department of Physics, Shanghai Jiao Tong University, Shanghai 200240, China}
\author{Lie-Wen Chen\footnote{%
Corresponding author (email: lwchen$@$sjtu.edu.cn)}}
\affiliation{Department of Physics, Shanghai Jiao Tong University, Shanghai 200240, China}
\affiliation{Center of Theoretical Nuclear Physics, National Laboratory of Heavy Ion
Accelerator, Lanzhou 730000, China}
\author{Bao-An Li}
\affiliation{Department of Physics and Astronomy, Texas A$\&$M University-Commerce,
Commerce, Texas 75429-3011, USA}
\author{Xiao-Hua Li}
\affiliation{Department of Physics, Shanghai Jiao Tong University, Shanghai 200240, China}
\affiliation{School of Nuclear Science and Technology, University of South China,
Hengyang, Hunan 421001, China}
\author{Chang Xu}
\affiliation{Department of Physics, Nanjing University, Nanjing 210008, China}
\date{\today}

\begin{abstract}
The nuclear symmetry energy $E_{sym}(\rho )$ and its density slope $L(\rho )$
can be decomposed analytically in terms of the single-nucleon potential in
isospin asymmetric nuclear matter. Using three popular nuclear effective
interaction models which have been extensively used in nuclear structure and
reaction studies, namely, the isospin and momentum dependent MDI interaction
model, the Skyrme Hartree-Fock approach and the Gogny Hartree-Fock approach,
we analyze the contribution of different terms in the single-nucleon
potential to the $E_{sym}(\rho )$ and $L(\rho )$. Our results show that the
observed different density behaviors of $E_{sym}(\rho )$ for different
interactions are essentially due to the variation of the symmetry potential $%
U_{sym,1}(\rho ,k)$. Furthermore, we find that the contribution of the
second-order symmetry potential $U_{sym,2}(\rho ,k)$ to the $L(\rho )$
generally cannot be neglected. Moreover, our results demonstrate that the
magnitude of the $U_{sym,2}(\rho ,k)$ is generally comparable with that of $%
U_{sym,1}(\rho ,k)$, indicating the second-order symmetry potential $%
U_{sym,2}(\rho ,k)$ may have significant corrections to the widely used Lane
approximation to the single-nucleon potential in extremely
neutron(proton)-rich nuclear matter.
\end{abstract}

\maketitle

\section{Introduction}

During the last decade, the nuclear symmetry energy $E_{sym}(\rho )$ that
essentially characterizes the isospin dependent part of the equation of
state (EOS) of asymmetric nuclear matter has attracted much attention from
different fields due to its multifaceted influences in nuclear physics and
astrophysics \cite{LiBA98,Dan02,Lat04,Ste05,Bar05,LCK08} as well as some
interesting issues regarding possible new physics beyond the standard model
\cite{Hor01b,Sil05,Kra07,Wen09}. For example, the density slope $L$ of the
symmetry energy at nuclear matter saturation density $\rho _{0}$ has been
shown to be important in determining several critical quantities such as the
size of neutron-skin in heavy nuclei \cite%
{Bro00,Hor01,Fur02,Die03,Che05b,Tod05,Cen09,Sam09,Che10,Roc11}, location of
neutron dripline \cite{Kaz10}, core-crust transition density \cite%
{Hor01,Lat04,Ste05,Oya07,Kub07,XuJ09,New11} and gravitational binding energy
\cite{New09} of neutron stars. The symmetry energy may also have significant
influence on gravitational wave emission from compact stars \cite%
{Kra08,Wen09b,Lin11,Gea11,Wen11b}. Furthermore, knowledge on the symmetry
energy might be useful to understand the non-Newtonian gravity proposed in
grand unified theories and to constrain properties of the neutral weakly
coupled light spin-1 gauge $U$-boson originated from supersymmetric
extensions of the standard model \cite{Kri09,Wen09,Zha11,Wen11,Zhe11}.

In recent years, a great deal of experimental and theoretical efforts have
been devoted to determining the density dependence of the symmetry energy
\cite{Bar05,LCK08}. Although significant progress has been made, large
uncertainties on $E_{sym}(\rho )$ still exist even around nuclear matter
saturation density, e.g., while the value of $E_{sym}(\rho _{0})$ is
determined to be around $30\pm 4$ MeV, mostly from analyzing nuclear masses,
the extracted density slope $L$ scatters in a very large range from about $%
20 $ to $115$ MeV, depending on the observables and methods used in the
studies \cite{Che05,Tsa09,Cen09} (See, e.g., Refs. \cite%
{XuC10,Che10,Tsa11,Che11a,New11b} for a review of recent progress). To
reduce the uncertainties on the constraints of $E_{sym}(\rho _{0})$ and $L$
is thus of critical importance and remains a big challenge in the community.

So far, information on $E_{sym}(\rho _{0})$ and its density slope $L$ is
essentially obtained from theoretical model analysis on the experimental
data of heavy ion collisions \cite{Tsa04,Che05,Fam06,She07,Tsa09}, nuclear
mass \cite{Mye96,Liu10}, excitation energies of isobaric analog states \cite%
{Dan09}, pygmy dipole resonance of neutron-rich nuclei \cite{Kli07,Car10},
isovector giant dipole resonance of neutron-rich nuclei \cite{Tri08,Cao08},
and neutron skin thickness \cite{Cen09,Che10}. In these theoretical models,
an energy density functional with a number of parameters is usually assumed
apriori, and the model parameters are then obtained from fitting
experimental data and the empirical values of some physical quantities.
Information on $E_{sym}(\rho _{0})$ and $L$ is then extracted based on the
obtained model parameters. Since all the phenomena/observables are in
someway at least indirectly and qualitatively related to the $E_{sym}(\rho
_{0})$ and $L$, it is very useful to directly decompose $E_{sym}(\rho _{0})$
and $L$ in terms of some relevant parts of the commonly used underlying
nuclear effective interaction \cite{XuC11}. This decomposition of the $%
E_{sym}(\rho _{0})$ and $L$ provides an important and physically more
transparent approach to extract information about isospin dependence of
strong interaction in nuclear medium from experiments and understand why the
predicted symmetry energy from various models is so uncertain \cite{XuC10b}.

In a recent work \cite{XuC10}, based on the Hugenholtz--Van Hove (HVH)
theorem \cite{Hug58,Sat99}, it was indeed shown that both $E_{sym}(\rho
_{0}) $ and $L$ are completely determined by the single-nucleon potential in
asymmetric nuclear matter which can be extracted from the nucleon global
optical model potentials. In that work, the Lane approximation \cite{Lan62}
to the single-nucleon potential in asymmetric nuclear matter, i.e., $%
U_{n/p}(\rho ,\delta ,k)\approx U_{0}(\rho ,k)\pm U_{sym,1}(\rho ,k)\delta $%
, has\ been assumed and also the momentum independent nucleon effective mass
has been used, therefore, the higher-order effects such as the contributions
from the second-order symmetry potential $U_{sym,2}(\rho ,k)$ and the
momentum dependence of the nucleon effective mass\ which can also contribute
to $L$ have been neglected. So far, to our best knowledge, there is not any
empirical or experimental information or even any theoretical predictions
on\ the second-order symmetry potential $U_{sym,2}(\rho ,k)$. It is thus
interesting and important to estimate the $U_{sym,2}(\rho ,k)$ with some
well-established theoretical models.

The main motivation of the present work is to evaluate the $U_{sym,2}(\rho
,k)$ and estimate its contribution to $L$ based on several popular nuclear
effective interaction models which have been extensively used in nuclear
structure and reaction studies. Our results indicate that, although the
momentum dependence of the nucleon effective mass might not be important,
the second-order symmetry potential $U_{sym,2}(\rho ,k)$\ might have
nonnegligible contribution to the $L$. Furthermore, we find that the
magnitude of the $U_{sym,2}(\rho ,k)$ is generally comparable with that of $%
U_{sym,1}(\rho ,k)$, indicating the second-order symmetry potential $%
U_{sym,2}(\rho ,k)$ may have significant corrections to the widely used Lane
approximation $U_{n/p}(\rho ,\delta ,k)\approx U_{0}(\rho ,k)\pm
U_{sym,1}(\rho ,k)\delta $ for the single-nucleon potential in extremely
neutron(proton)-rich nuclear matter, e.g., in neutron stars and the
neutron-skin region of heavy nuclei. These results imply that it is
important to extract experimentally information on $U_{sym,2}(\rho ,k)$.

The paper is organized as follows. In Section \ref{theory}, we briefly
recall the definition of the symmetry energy and the symmetry potential in
asymmetric nuclear matter, and then derive the explicit expressions for the
single-nucleon potential decomposition of the symmetry energy and its
density slope. The results and discussions are presented in Section \ref%
{results}. A summary is then given in Section \ref{summary}. For
completeness, the theoretical models used in the present paper are briefly
described in the Appendix.

\section{Theoretical formulism}

\label{theory}

\subsection{The symmetry energy and the symmetry potential in asymmetric
nuclear matter\label{secfundation}}

Due to the exchange symmetry between protons and neutrons in nuclear matter
when one neglects the Coulomb interaction and assumes the charge symmetry of
nuclear forces, the EOS of isospin asymmetric nuclear matter, defined by its
binding energy per nucleon, can be expanded as a power series of even-order
terms in isospin asymmetry $\delta $ as
\begin{equation}
E(\rho ,\delta )=E_{0}(\rho )+E_{\mathrm{sym}}(\rho )\delta ^{2}+O(\delta
^{4}),  \label{EOSANM}
\end{equation}%
where $\rho =\rho _{n}+\rho _{p}$ is the baryon density and $\delta =(\rho
_{n}-\rho _{p})/\rho $ is the isospin asymmetry with $\rho _{n}$ and $\rho
_{p}$ denoting the neutron and proton densities, respectively; $E_{0}(\rho
)=E(\rho ,\delta =0)$ is the EOS of symmetric nuclear matter, and the
nuclear symmetry energy is expressed as
\begin{equation}
E_{\mathrm{sym}}(\rho )=\frac{1}{2!}\frac{\partial ^{2}E(\rho ,\delta )}{%
\partial \delta ^{2}}|_{\delta =0}.  \label{Esym}
\end{equation}%
The higher-order terms of $\delta $ in Eq. (\ref{EOSANM}) are negligible,
leading to the well-known empirical parabolic law for the EOS of asymmetric
nuclear matter, which has been verified by all many-body theories to date,
at least for densities up to moderate values (See, e.g., Ref. \cite{LCK08}).

Around the nuclear matter saturation density $\rho _{0}$, the nuclear
symmetry energy $E_{\mathrm{sym}}(\rho )$\ can be expanded as
\begin{equation}
E_{\mathrm{sym}}(\rho )=E_{\text{\textrm{sym}}}({\rho _{0}})+L\chi +O(\chi
^{2}),
\end{equation}%
where $\chi =(\rho -\rho _{0})/3\rho _{0}$ is a dimensionless variable and $%
L=L(\rho _{0})$ is the density slope parameter of the symmetry energy at $%
\rho _{0}$. More generally, the slope parameter of the symmetry energy at
arbitrary density $\rho $ is defined as
\begin{equation}
L(\rho )=3\rho \frac{dE_{\mathrm{sym}}(\rho )}{d\rho }.  \label{L}
\end{equation}%
The slope parameter $L$ at $\rho _{0}$ characterizes the density dependence
of the nuclear symmetry energy around nuclear matter saturation density $%
\rho _{0}$, and thus carry important information on the properties of
nuclear symmetry energy at both high and low densities.

The single-nucleon potential $U_{\tau }(\rho ,\delta ,k)$ (we assume $\tau
=1 $ for neutrons and $-1$ for protons in this work) in asymmetric nuclear
matter generally depends on the baryon density $\rho $, the isospin
asymmetry $\delta $ and the amplitude of the nucleon momentum $k$. Due to
the isospin symmetry of nuclear interactions under the exchange of neutrons
and protons, the single-nucleon potential $U_{\tau }(\rho ,\delta ,k)$ can
be expanded as a power series of $\delta $ as \cite{XuC11}%
\begin{eqnarray}
U_{\tau }(\rho ,\delta ,k) &=&U_{0}(\rho ,k)+\sum_{i=1,2,\cdot \cdot \cdot
}U_{sym,i}(\rho ,k)(\tau \delta )^{i}  \notag \\
&=&U_{0}(\rho ,k)+U_{sym,1}(\rho ,k)(\tau \delta )  \notag \\
&&+U_{sym,2}(\rho ,k)(\tau \delta )^{2}+\cdot \cdot \cdot ,
\label{UtauTaylor}
\end{eqnarray}%
where $U_{0}(\rho ,k)\equiv U_{n}(\rho ,0,k)=U_{p}(\rho ,0,k)$ is the
single-nucleon potential in symmetric nuclear matter and $U_{sym,i}(\rho ,k)$
are expressed as
\begin{eqnarray}
U_{sym,i}(\rho ,k) &\equiv &\frac{1}{i!}\frac{\partial ^{i}U_{n}(\rho
,\delta ,k)}{\partial \delta ^{i}}|_{\delta =0}  \notag \\
&=&\frac{(-1)^{i}}{i!}\frac{\partial ^{i}U_{p}(\rho ,\delta ,k)}{\partial
\delta ^{i}}|_{\delta =0},  \label{defusymn}
\end{eqnarray}%
with $U_{sym,1}(\rho ,k)$ being the well-known nuclear symmetry potential
\cite{LCK08} (where $U_{sym,1}$ is denoted by $U_{sym}$), and\ the
higher-order term $U_{sym,2}(\rho ,k)$ being called as the second-order
nuclear symmetry potential here. Neglecting the higher-order terms ($\delta
^{2}$, $\delta ^{3}$, $\cdot \cdot \cdot $) in Eq. (\ref{UtauTaylor}) leads
to the well-known Lane potential \cite{Lan62}, i.e.,%
\begin{equation}
U_{\tau }(\rho ,\delta ,k)\approx U_{0}(\rho ,k)+U_{sym,1}(\rho ,k)(\tau
\delta ),  \label{Lane}
\end{equation}%
which has been extensively used to approximate the single-nucleon potential $%
U_{\tau }(\rho ,\delta ,k)$ in asymmetric nuclear matter, and in this case
the symmetry potential $U_{sym,1}(\rho ,k)$ can be obtained approximately by
\cite{LCK08,LiBA04b}
\begin{equation}
U_{sym,1}(\rho ,k)\approx \frac{U_{n}(\rho ,\delta ,k)-U_{p}(\rho ,\delta ,k)%
}{2\delta }.  \label{Usym1Lane}
\end{equation}

\subsection{Single-nucleon potential decomposition of the symmetry energy
and its density slope}

\label{secdecompo}

According to the HVH theorem \cite{Hug58,Sat99}, the chemical potentials of
neutrons and protons in asymmetric nuclear matter with energy density $%
\varepsilon (\rho ,\delta )=\rho E(\rho ,\delta )$ can be expressed,
respectively, as
\begin{eqnarray}
t(k_{F_{n}})+U_{n}(\rho ,\delta ,k_{F_{n}}) &=&\frac{\partial \varepsilon
(\rho ,\delta )}{\partial \rho _{n}},  \label{chemUn} \\
t(k_{F_{p}})+U_{p}(\rho ,\delta ,k_{F_{p}}) &=&\frac{\partial \varepsilon
(\rho ,\delta )}{\partial \rho _{p}},  \label{chemUp}
\end{eqnarray}%
where $t(k_{F_{\tau }})=k_{F_{\tau }}^{2}/2m$ is the nucleon kinetic energy
at Fermi momentum $k_{F_{\tau }}=k_{F}(1+\tau \delta )^{1/3}$ with $%
k_{F}=(3\pi ^{2}\rho /2)^{1/3}$ being the Fermi momentum in symmetric
nuclear matter at density $\rho $. We would like to point out that the HVH
theorem is independent of the detailed nature of the nucleon interactions
used and has been strictly proven to be valid for any interacting self-bound
infinite Fermi system \cite{Hug58,Sat99}.

The right-hand side of Eq. (\ref{chemUn}) can be further written as
\begin{eqnarray}
\frac{\partial \varepsilon (\rho ,\delta )}{\partial \rho _{n}} &=&\frac{%
\partial \varepsilon (\rho ,\delta )}{\partial \rho }\frac{\partial \rho }{%
\partial \rho _{n}}+\frac{\partial \varepsilon (\rho ,\delta )}{\partial
\delta }\frac{\partial \delta }{\partial \rho _{n}}  \notag \\
&=&\frac{\partial \varepsilon (\rho ,\delta )}{\partial \rho }+\frac{1}{\rho
}\frac{\partial \varepsilon (\rho ,\delta )}{\partial \delta }-\frac{%
\partial \varepsilon (\rho ,\delta )}{\partial \delta }\frac{\delta }{\rho }.
\label{chemUn1}
\end{eqnarray}%
Similarly, the right-hand side of Eq. (\ref{chemUp}) can be expressed as%
\begin{equation}
\frac{\partial \varepsilon (\rho ,\delta )}{\partial \rho _{p}}=\frac{%
\partial \varepsilon (\rho ,\delta )}{\partial \rho }-\frac{1}{\rho }\frac{%
\partial \varepsilon (\rho ,\delta )}{\partial \delta }-\frac{\partial
\varepsilon (\rho ,\delta )}{\partial \delta }\frac{\delta }{\rho }.
\label{chemUp1}
\end{equation}%
Subtracting Eq. (\ref{chemUp1}) from Eq. (\ref{chemUn1}) and noting $%
\varepsilon (\rho ,\delta )=\rho E(\rho ,\delta )$, we then obtain
\begin{eqnarray}
&&\frac{\partial \varepsilon (\rho ,\delta )}{\partial \rho _{n}}-\frac{%
\partial \varepsilon (\rho ,\delta )}{\partial \rho _{p}}  \notag \\
&=&\frac{2}{\rho }\frac{\partial \varepsilon (\rho ,\delta )}{\partial
\delta }=2\frac{\partial E(\rho ,\delta )}{\partial \delta },
\label{chemnpminus}
\end{eqnarray}%
while adding Eq. (\ref{chemUn1}) and Eq. (\ref{chemUp1}), we have%
\begin{eqnarray}
&&\frac{\partial \varepsilon (\rho ,\delta )}{\partial \rho _{n}}+\frac{%
\partial \varepsilon (\rho ,\delta )}{\partial \rho _{p}}  \notag \\
&=&2E(\rho ,\delta )+2\rho \frac{\partial E(\rho ,\delta )}{\partial \rho }%
-2\delta \frac{\partial E(\rho ,\delta )}{\partial \delta }.
\label{chemnpplus}
\end{eqnarray}%
On the one hand, substituting Eq. (\ref{EOSANM}) into Eq. (\ref{chemnpminus}%
) and Eq. (\ref{chemnpplus}) respectively leads to following expressions,
i.e.,
\begin{eqnarray}
&&t(k_{F_{n}})-t(k_{F_{p}})+U_{n}(\rho ,\delta ,k_{F_{n}})-U_{p}(\rho
,\delta ,k_{F_{p}})  \notag \\
&=&4E_{sym}(\rho )\delta +\mathcal{O}(\delta ^{3})  \label{tayloreme}
\end{eqnarray}%
and%
\begin{eqnarray}
&&t(k_{F_{n}})+t(k_{F_{p}})+U_{n}(\rho ,\delta ,k_{F_{n}})+U_{p}(\rho
,\delta ,k_{F_{p}})  \notag \\
&=&2E_{0}(\rho )+2\rho \frac{\partial E_{0}(\rho )}{\partial \rho }  \notag
\\
&+&\Big[\frac{2}{3}L(\rho )-2E_{sym}(\rho )\Big]\delta ^{2}  \notag \\
&+&\mathcal{O}(\delta ^{4}).  \label{taylorepe}
\end{eqnarray}%
On the other hand, $t(k_{F_{\tau }})$ and $U_{\tau }(\rho ,\delta
,k_{F_{\tau }})$ can be expanded as a power series of $\delta $,
respectively, as
\begin{eqnarray}
&&t(k_{F_{\tau }})=t(k_{F})  \notag \\
&+&\frac{\partial t(k)}{\partial k}|_{k_{F}}\cdot \frac{1}{3}k_{F}(\tau
\delta )  \notag \\
&+&\frac{1}{2}\Big[\frac{k_{F}^{2}}{9}{\frac{\partial ^{2}t(k)}{\partial
k^{2}}}|_{k_{F}}-\frac{2k_{F}}{9}{\frac{\partial t(k)}{\partial k}}|_{k_{F}}%
\Big]\delta ^{2}  \notag \\
&+&\mathcal{O}(\delta ^{3}),  \label{taylort}
\end{eqnarray}%
and%
\begin{eqnarray}
&&U_{\tau }(\rho ,\delta ,k_{F_{\tau }})=U_{0}(\rho ,k_{F})  \notag \\
&+&\Big[\frac{k_{F}}{3}{\frac{\partial U_{0}(\rho ,k)}{\partial k}}%
|_{k_{F}}+U_{sym,1}(\rho ,k_{F})\Big](\tau \delta )  \notag \\
&+&\Big[\frac{k_{F}}{3}{\frac{\partial U_{sym,1}(\rho ,k)}{\partial k}}%
|_{k_{F}}+U_{sym,2}(\rho ,k_{F})\Big]\delta ^{2}  \notag \\
&+&\frac{1}{2}\Big[\frac{k_{F}^{2}}{9}{\frac{\partial ^{2}U_{0}(\rho ,k)}{%
\partial k^{2}}}|_{k_{F}}-\frac{2k_{F}}{9}{\frac{\partial U_{0}(\rho ,k)}{%
\partial k}}|_{k_{F}}\Big]\delta ^{2}  \notag \\
&+&\mathcal{O}(\delta ^{3}).  \label{taylorU}
\end{eqnarray}%
Substituting Eqs. (\ref{taylort}) and (\ref{taylorU}) into the left-hand
sides of Eqs. (\ref{tayloreme}) and (\ref{taylorepe}), and comparing the
coefficients of\ the $1$st-order $\delta $ terms in both left and right hand
sides, we then obtain
\begin{equation}
E_{sym}(\rho )=\frac{1}{2}U_{sym,1}(\rho ,k_{F})+\frac{1}{6}\frac{\partial
\lbrack t(k)+U_{0}(\rho ,k)]}{\partial k}|_{k_{F}}\cdot k_{F},
\label{Esymexp}
\end{equation}%
while comparing the coefficients of $2$nd-order $\delta $ terms in both
sides leads to the following expression
\begin{eqnarray}
L(\rho ) &=&\frac{3}{2}U_{sym,1}(\rho ,k_{F})+3U_{sym,2}(\rho ,k_{F})  \notag
\\
&+&\frac{\partial U_{sym,1}}{\partial k}|_{k_{F}}\cdot k_{F}+\frac{1}{6}%
\frac{\partial \lbrack t(k)+U_{0}(\rho ,\delta )]}{\partial k}|_{k_{F}}\cdot
k_{F}  \notag \\
&+&\frac{1}{6}\frac{\partial ^{2}[t(k)+U_{0}(\rho ,\delta )]}{\partial k^{2}}%
|_{k_{F}}\cdot k_{F}^{2}.  \label{Lexp}
\end{eqnarray}%
It should be stressed that higher-order terms of the single-nucleon
potential in Eq. (\ref{UtauTaylor}) (i.e., $U_{sym,3}(\rho ,k)$ and
higher-order terms) have no contributions to $E_{sym}(\rho )$ and $L(\rho )$%
, and thus Eq. (\ref{Esymexp}) and Eq. (\ref{Lexp}) are complete and exact,
and are valid for arbitrary density $\rho $. Furthermore, Eq. (\ref{Esymexp}%
) and Eq. (\ref{Lexp}) can be rewritten as
\begin{widetext}
\begin{eqnarray}
E_{sym}(\rho)&=& \frac{1}{3}\frac{\hbar^2 k^2}{2m_0^*}|_{k_F}
+ \frac{1}{2} U_{sym,1}(\rho,k_F),  \label{Esymexpfinal}
\\
L(\rho) &=& \frac{2}{3}\frac{\hbar^2 k^2}{2m_0^*}|_{k_F}
- \frac{1}{6}\Big(\frac{\hbar^2 k^3}{{m_0^*}^2}\frac{\partial m_0^*}{\partial k} \Big)|_{k_F}
  + \frac{3}{2}U_{sym,1}(\rho,k_F) + \frac{\partial U_{sym,1}}{\partial k}|_{k_F}\cdot k_F
  + 3U_{sym,2}(\rho,k_F),   \label{Lexpfinal}
\end{eqnarray}
\end{widetext}in terms of the nucleon effective mass $m_{0}^{\ast }(\rho ,k)$
in symmetric nuclear matter which is generally dependent of the density $%
\rho $ and the nucleon momentum $k$, i.e.,%
\begin{equation}
m_{0}^{\ast }(\rho ,k)={\ \frac{m}{1+\frac{m}{\hbar ^{2}k}\frac{\partial
U_{0}(\rho ,k)}{\partial k}},}  \notag  \label{m0eff}
\end{equation}%
due to the following relations
\begin{eqnarray}
\frac{\partial \lbrack t(k)+U_{0}(\rho ,k)]}{\partial k}|_{k_{F}} &=&\frac{%
\hbar ^{2}k}{m_{0}^{\ast }}|_{k_{F}},  \notag \\
\frac{\partial ^{2}[t(k)+U_{0}(\rho ,\delta )]}{\partial k^{2}}|_{k_{F}} &=&%
\frac{\hbar ^{2}}{m_{0}^{\ast }}|_{k_{F}}-\Big(\frac{\hbar ^{2}k}{{%
m_{0}^{\ast }}^{2}}\frac{\partial m_{0}^{\ast }}{\partial k}\Big)|_{k_{F}}.
\notag
\end{eqnarray}

For convenience, we can reexpress $E_{sym}(\rho )$ and $L(\rho )$,
respectively, as
\begin{eqnarray}  \label{L12345}
E_{sym}(\rho ) &=&E_{1}(\rho )+E_{2}(\rho ),  \label{e1e2} \\
L(\rho ) &=&L_{1}(\rho )+L_{2}(\rho )+L_{3}(\rho )+L_{4}(\rho )+L_{5}(\rho ),
\notag
\end{eqnarray}%
with
\begin{eqnarray}
E_{1}(\rho ) &=&\frac{1}{3}\frac{\hbar ^{2}k_{F}^{2}}{2m_{0}^{\ast }(\rho
,k_{F})} \\
E_{2}(\rho ) &=&\frac{1}{2}U_{sym,1}(\rho ,k_{F}) \\
L_{1}(\rho ) &=&\frac{2}{3}\frac{\hbar ^{2}k_{F}^{2}}{2m_{0}^{\ast }(\rho
,k_{F})} \\
L_{2}(\rho ) &=&-\frac{1}{6}\frac{\hbar ^{2}k_{F}^{3}}{{m_{0}^{\ast }}%
^{2}(\rho ,k_{F})}\frac{\partial m_{0}^{\ast }(\rho ,k)}{\partial k}|_{k_{F}}
\\
L_{3}(\rho ) &=&\frac{3}{2}U_{sym,1}(\rho ,k_{F}) \\
L_{4}(\rho ) &=&\frac{\partial U_{sym,1}(\rho ,k)}{\partial k}|_{k_{F}}\cdot
k_{F} \\
L_{5}(\rho ) &=&3U_{sym,2}(\rho ,k_{F}).
\end{eqnarray}%
In this way, one can see that $E_{1}(\rho )$ represents the kinetic energy
part (including the effective mass contribution) of the symmetry energy
while $E_{2}(\rho )$ is due to the symmetry potential contribution to the
symmetry energy. Furthermore, $L_{1}(\rho )$, $L_{2}(\rho )$, $L_{3}(\rho )$%
, $L_{4}(\rho )$, and $L_{5}(\rho )$ have respective physical meaning,
namely, $L_{1}(\rho )$ represents the kinetic energy part (including the
effective mass contribution) of the $L$ parameter, $L_{2}(\rho )$ is from
the momentum dependence of the nucleon effective mass, $L_{3}(\rho )$ is due
to the symmetry potential contribution, $L_{4}(\rho )$ comes from the
momentum dependence of the symmetry potential, and $L_{5}(\rho )$ is from
the second-order symmetry potential $U_{sym,2}(\rho ,k_{F})$. In this way,
the symmetry energy $E_{sym}(\rho )$ and its slope $L(\rho )$ have been
decomposed in terms of $U_{0}(\rho ,k)$, $U_{sym,1}(\rho ,k)$, $%
U_{sym,2}(\rho ,k)$ and/or their $1$st and $2$nd-order partial derivatives
with respect to $k$ and $\delta $. In particular, at nuclear matter
saturation density $\rho _{0}$, $U_{0}(\rho _{0},k)$, $U_{sym,1}(\rho
_{0},k) $ and $U_{sym,2}(\rho _{0},k)$ (and thus $E_{sym}(\rho _{0})$ and
its slope parameter $L$) can be determined completely from the\ isospin and
momentum dependent nucleon global optical potential, which can be directly
extracted from nucleon-nucleus and (p,n) charge-exchange reactions (See,
e.g., Refs. \cite{XuC10,Kon03,LiXH11}).

For the single-nucleon potential decomposition of the slope parameter $%
L(\rho )$, if we use the Lane potential Eq. (\ref{Lane}) and neglect the
contributions from the momentum dependence of the nucleon effective mass,
the $L(\rho )$ is then reduced to
\begin{equation}
L(\rho )=L_{1}+L_{3}+L_{4},
\end{equation}%
namely,%
\begin{eqnarray}
L(\rho ) &=&\frac{2}{3}\frac{\hbar ^{2}k^{2}}{2m_{0}^{\ast }}|_{k_{F}}+\frac{%
3}{2}U_{sym,1}(\rho ,k_{F})  \notag \\
&+&\frac{\partial U_{sym,1}(\rho ,k)}{\partial k}|_{k_{F}}\cdot k_{F},
\label{XCL}
\end{eqnarray}%
which has been used in previous work \cite{XuC10}. Although the Lane
potential could be a good approximation in evaluating $U_{sym,1}(\rho ,k)$
as in Eq. (\ref{Usym1Lane}) \cite{LCK08,LiBA04b}, it would be interesting to
see if the higher-order $U_{sym,2}(\rho ,k_{F})$ contribution to\ the $%
L(\rho )$ is significant or not. Using three popular nuclear effective
interaction models, we will demonstrate in the following that the
contribution of higher-order $U_{sym,2}(\rho ,k)$ term\ to $L(\rho )$
generally cannot be neglected.

\section{Results and Discussions}

\label{results}

In the following, we analyze the single-nucleon potential decomposition of $%
E_{sym}(\rho )$ and $L(\rho )$ as well as the density and momentum
dependence of $U_{sym,1}(\rho ,k)$ and $U_{sym,2}(\rho ,k)$ using three
popular nuclear effective interaction models which have been extensively
used in nuclear structure and reaction studies, namely, the isospin and
momentum dependent MDI interaction model, the Skyrme Hartree-Fock approach,
and the Gogny Hartree-Fock approach. One can find the details about these
three models in the Appendix. A very useful feature of these models is that
analytical expressions for many interesting physical quantities, such as the
single-nucleon potential in asymmetric nuclear matter at zero temperature,
can be obtained, and this makes our analysis and calculations physically
transparent and very convenient.

For the MDI interaction, we use $3$ parameter sets, i.e., $x=-1$, $x=0$ and $%
x=1$ \cite{Che05}, which\ give three different density dependence of the
symmetry energy, namely, stiff one, moderate one, and soft one,
respectively, and have been applied extensively in transport model
simulations for heavy ion collisions. For the Skyrme interaction, we mainly
use the famous SKM* \cite{Bart82a} and SLy4 \cite{Chaba98a} as well as the
recently developed MSL0 \cite{Che10}. In addition, a number of other Skyrme
interactions are used for the single-nucleon potential decomposition of $%
E_{sym}(\rho )$ and $L(\rho )$ at $\rho _{0}$. For the Gogny interaction, we
use the existing D1 \cite{Dec80}, D1S \cite{Berger91a}, D1N \cite{Chap08a},
and D1M \cite{Gor09a} which have been successfully applied in nuclear
structure studies.

\subsection{Single-nucleon potential decomposition of $E_{sym}(\protect\rho %
) $}

\begin{figure}[tbh]
\centering
\includegraphics[width=8.6cm]{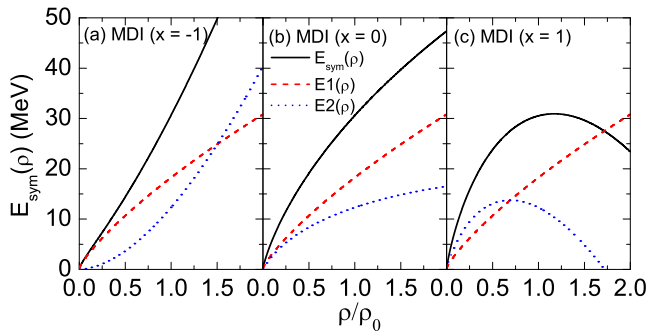}
\caption{(Color online) Density dependence of $E_{sym}(\protect\rho )$, $%
E_{1}(\protect\rho )=\frac{1}{3}\frac{\hbar ^{2}k_{F}^{2}}{2m_{0}^{\ast }(%
\protect\rho ,k_{F})}$ and $E_{2}(\protect\rho )=\frac{1}{2}U_{sym,1}(%
\protect\rho ,k_{F})$ in the MDI interaction with $x=-1$ (a), $0$ (b), and $%
1 $ (c).}
\label{EsymMDI}
\end{figure}

\begin{figure}[tbh]
\centering
\includegraphics[width=8.6cm]{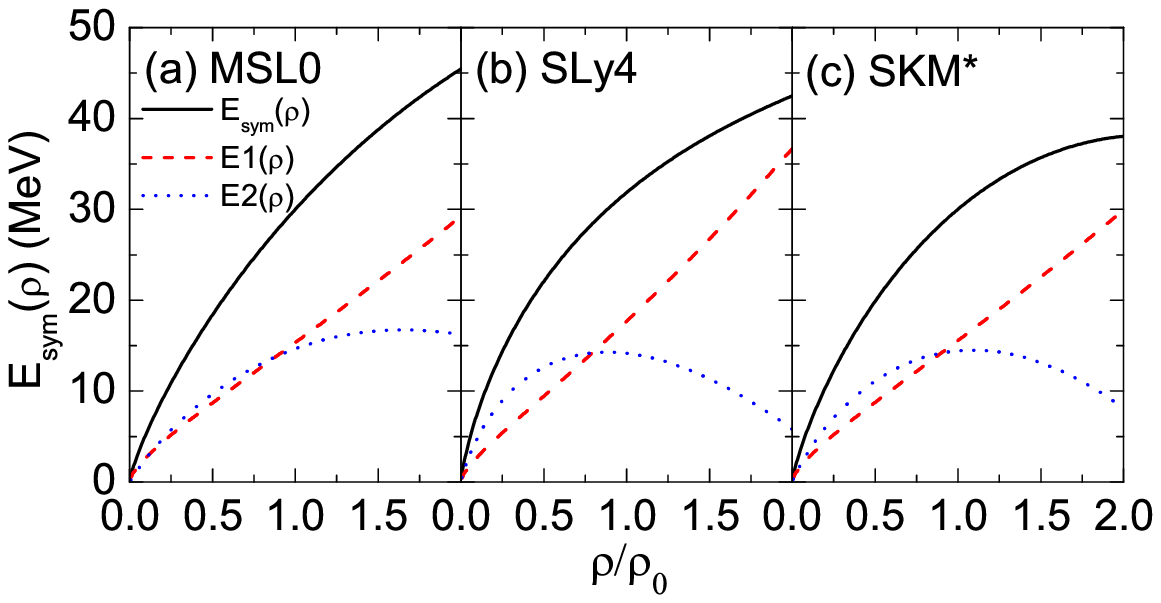}
\caption{(Color online) Same as Fig. \protect\ref{EsymMDI} but in the Skyrme
Hartree-Fock\ approach with MSL0 (a), SLy4 (b), and SKM* (c).}
\label{EsymSHF}
\end{figure}

\begin{figure}[tbh]
\centering
\includegraphics[width=8.5cm]{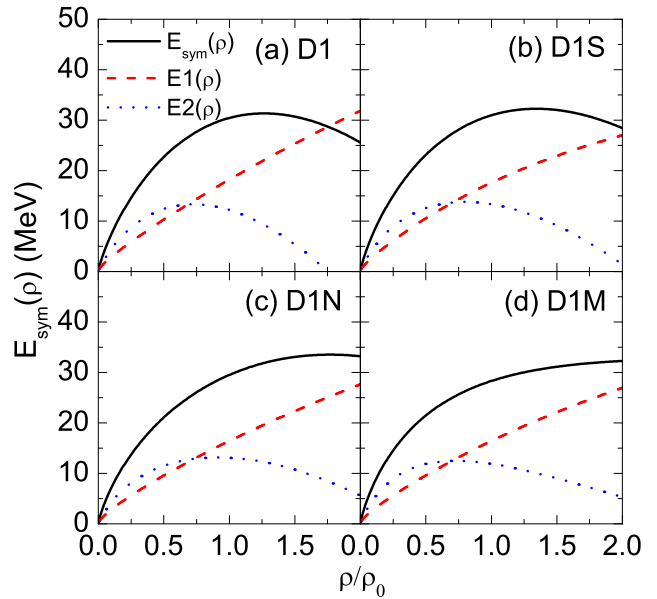}
\caption{(Color online) Same as Fig. \protect\ref{EsymMDI} but in Gogny
Hartree-Fock\ approach with D1 (a), D1S (b), D1N (c), and D1M (d).}
\label{EsymGogny}
\end{figure}

In Fig. \ref{EsymMDI}, Fig. \ref{EsymSHF} and Fig. \ref{EsymGogny}, we plot
the density dependence of $E_{sym}(\rho )$, $E_{1}(\rho )$ and $E_{2}(\rho )$
in the MDI interaction model, the Skyrme-Hartree-Fock approach, and the
Gogny-Hartree-Fock approach, respectively. One can see from Fig. \ref%
{EsymMDI} that for different $x$ values, the $E_{1}(\rho )$ displays the
same density dependence while the $E_{2}(\rho )$ exhibits very different
density behaviors, indicating that the different density dependences of $%
E_{sym}(\rho )$ for $x=-1$, $0$ and $1$ are completely due to the different
density dependence of $E_{2}(\rho )$, i.e., the symmetry potential $%
U_{sym,1}(\rho ,k_{F})$. The similar behaviors can also be seen from Fig. %
\ref{EsymSHF} for the Skyrme-Hartree-Fock calculations and Fig. \ref%
{EsymGogny} for the Gogny-Hartree-Fock calculations.

Furthermore, one can see from Fig. \ref{EsymMDI}, Fig. \ref{EsymSHF} and
Fig. \ref{EsymGogny} that for all the interactions considered here, the $%
E_{1}(\rho )$ increases with the density and is always positive while $%
E_{2}(\rho )$ can increase or decrease with the density and even become
negative at higher densities. These features can be understood since the $%
E_{1}(\rho )=\frac{1}{3}\frac{\hbar ^{2}k_{F}^{2}}{2m_{0}^{\ast }(\rho
,k_{F})}$ is determined uniquely by the single-nucleon potential $U_{0}(\rho
,k)$ in symmetric nuclear matter for which reliable information about its
density and momentum dependence has already been obtained from heavy-ion
collisions, see, e.g., Ref. \cite{Dan02}, albeit there is still some room
for further improvements, particularly at high momenta/densities, and the
nuclear effective interactions are usually constructed to describe
reasonably $U_{0}(\rho ,k)$, especially around $\rho _{0}$. However, on the
contrary, the symmetry potential $U_{sym,1}(\rho ,k_{F})$, which mainly
reflects the isospin dependence of the nuclear effective interaction in
nuclear medium, is still not very well determined, especially at high
densities and momenta. In fact, it has been identified as the key quantity
responsible for the uncertain high density behavior of the symmetry energy
as stressed in Ref. \cite{LCK08} (See also Refs. \cite{XuC10b,XuC11}). These
results show that the observed different density behaviors of $E_{sym}(\rho
) $ for different interactions are essentially due to the variation of the
symmetry potential $U_{sym,1}(\rho ,k)$.

\subsection{Single-nucleon potential decomposition of $L(\protect\rho )$}

\begin{figure}[tbh]
\includegraphics[width=8.6cm]{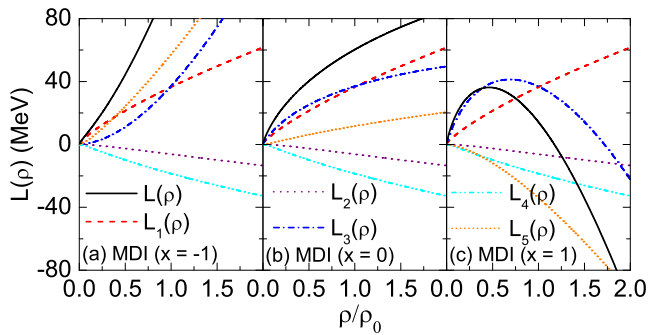}
\caption{(Color online) Density dependence of $L(\protect\rho )$, $L_{1}(%
\protect\rho )=\frac{2}{3}\frac{\hbar ^{2}k_{F}^{2}}{2m_{0}^{\ast }(\protect%
\rho ,k_{F})}$, $L_{2}(\protect\rho )=-\frac{1}{6}\frac{\hbar ^{2}k_{F}^{3}}{%
{m_{0}^{\ast }}^{2}(\protect\rho ,k_{F})}\frac{\partial m_{0}^{\ast }(%
\protect\rho ,k)}{\partial k}|_{k_{F}}$, $L_{3}(\protect\rho )=\frac{3}{2}%
U_{sym,1}(\protect\rho ,k_{F})$, $L_{4}(\protect\rho )=\frac{\partial
U_{sym,1}(\protect\rho ,k)}{\partial k}|_{k_{F}}\cdot k_{F}$, and $L_{5}(%
\protect\rho )=3U_{sym,2}(\protect\rho ,k_{F})$ in the MDI interaction with $%
x=-1$ (a), $0$ (b), and $1$ (c).}
\label{LMDI}
\end{figure}

\begin{figure}[tbh]
\includegraphics[width=8.6cm]{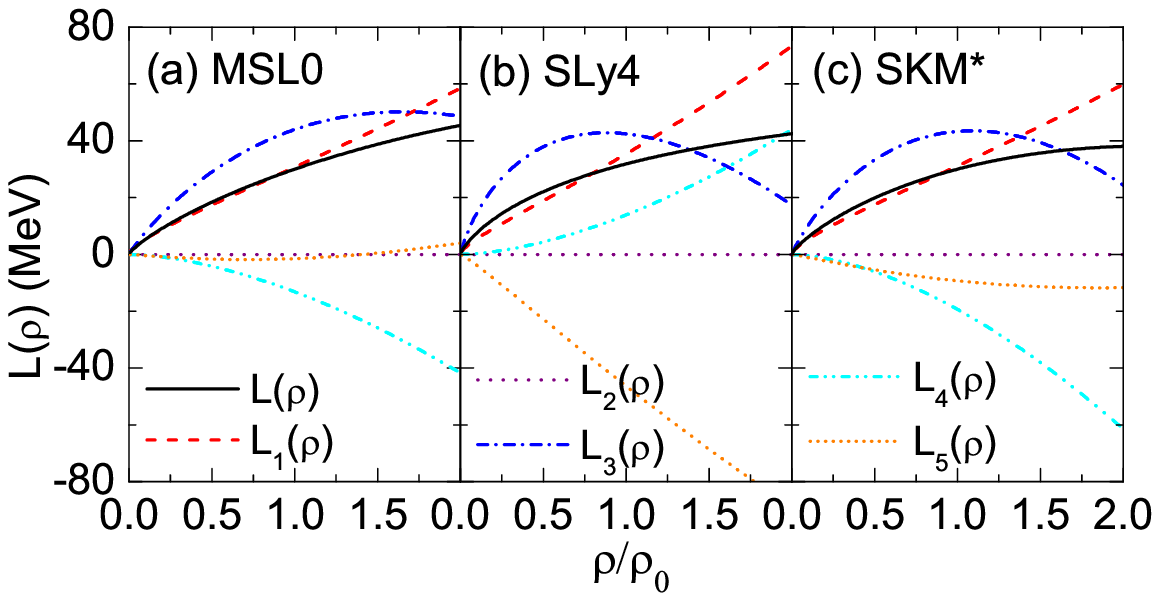}
\caption{(Color online) Same as Fig. \protect\ref{LMDI} but in the Skyrme
Hartree-Fock\ approach with MSL0 (a), SLy4 (b), and SKM* (c).}
\label{LSHF}
\end{figure}

\begin{figure}[tbh]
\includegraphics[width=8.5cm]{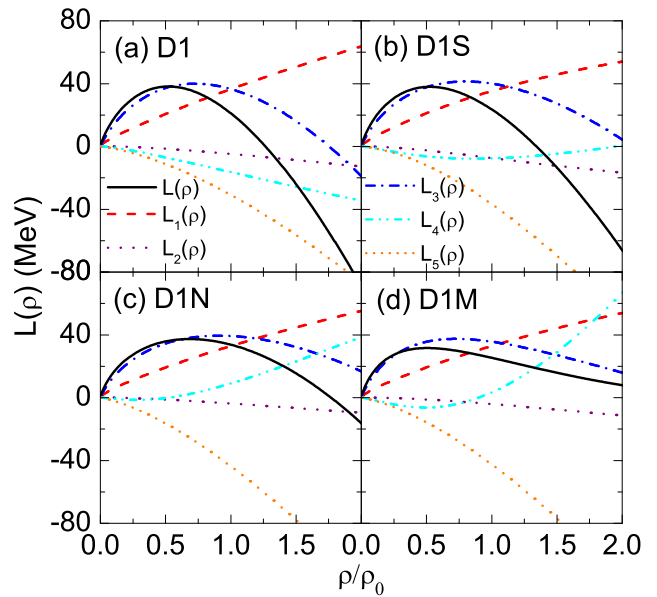}
\caption{(Color online) Same as Fig. \protect\ref{LMDI} but in Gogny
Hartree-Fock\ approach with D1 (a), D1S (b), D1N (c), and D1M (d).}
\label{LGogny}
\end{figure}

In order to illustrate the single-nucleon potential decomposition of $L(\rho
)$, we show in Fig. \ref{LMDI}, Fig. \ref{LSHF} and Fig. \ref{LGogny} the
density dependence of $L(\rho )$, $L_{1}(\rho )$, $L_{2}(\rho )$, $%
L_{3}(\rho )$, $L_{4}(\rho )$ and $L_{5}(\rho )$ in the MDI interaction
model, the Skyrme-Hartree-Fock approach, and the Gogny-Hartree-Fock
approach, respectively. It is seen that the $L_{1}(\rho )$\ displays the
similar density dependence for all the interactions considered here, just
like the $E_{1}(\rho )$ shown in Fig. \ref{EsymMDI}, Fig. \ref{EsymSHF} and
Fig. \ref{EsymGogny} because of $L_{1}(\rho )=2E_{1}(\rho )$. The $%
L_{2}(\rho )$ is seen to contribute a small negative value to the $L(\rho )$%
, indicating that the momentum dependence of the nucleon effective mass is
generally unimportant. In particular, one can see from Fig. \ref{LSHF} that $%
L_{2}(\rho )$ vanishes for the Skyrme-Hartree-Fock calculations due to the
fact that the nucleon effective mass is momentum independent for the
zero-range Skyrme interaction considered here. The $L_{3}(\rho )$ exhibits
different density dependences for different interactions, reflecting the
variation of $U_{sym,1}(\rho ,k_{F})$ with density for different
interactions. The $L_{4}(\rho )$ represents the contribution of the momentum
dependence of the symmetry potential to the $L(\rho )$, and it displays
different density dependence for different interactions and can be negative
or positive, depending on the interaction used.

It is particularly interesting to analyze the $L_{5}(\rho )$\ since it
reflects the higher-order $U_{sym,2}(\rho ,k_{F})$ contribution to the $%
L(\rho )$ and has been neglected in the previous work \cite{XuC10}. From
Fig. \ref{LMDI}, Fig. \ref{LSHF} and Fig. \ref{LGogny}, it is surprising to
see that the $L_{5}(\rho )$ may play an important role to determine the $%
L(\rho )$. In the MDI interaction with $x=-1$, the $L_{5}(\rho )$ is always
positive and increases rapidly with density while the opposite behavior is
observed for the MDI interaction with $x=1$. For the MDI interaction with $%
x=0$, the $L_{5}(\rho )$ is positive and moderately increases with density.
For the Skyrme-Hartree-Fock calculations, the $L_{5}(\rho )$ can be positive
or negative while it is always negative for the Gogny-Hartree-Fock
calculations with D1, D1S, D1N, and D1M. These results indicate that
generally the higher-order $U_{sym,2}(\rho ,k_{F})$ contribution to the $%
L(\rho )$ cannot be neglected and the Lane potential approximation
to the single-nucleon potential in asymmetric nuclear matter may
cause significant error for the determination of $L(\rho )$ from the
single-nucleon potential decomposition.

\begin{table*}[tbp]
\caption{The characteristic parameters $\protect\rho _{0}$ (fm$^{-3}$), $%
E_{0}(\protect\rho _{0})$ (MeV), $E_{sym}(\protect\rho _{0})$ (MeV), $E_{1}$
(MeV), $E_{2}$ (MeV), $L$ (MeV), $L_{1}$ (MeV), $L_{2}$ (MeV), $L_{3}$
(MeV), $L_{4}$ (MeV), and $L_{5}$ (MeV) at saturation density $\protect\rho %
_{0}$ for the MDI interaction with $x=-1$, $0$ and $1$, the
Gogny-Hartree-Fock predictions with D1, D1S, D1N and D1M as well as the
Skyrme-Hartree-Fock predictions with $112$ standard Skyrme interactions. The
interactions in different models are in order according to the $L$ value.
The corresponding reference is included as the last column.}
\label{TableI}%
\begin{tabular*}{\textwidth}{@{\extracolsep{\fill}}lcccccccccccr}
\toprule Model & $\rho_0$ & $E_0(\rho_0)$ & $E_{sym}(\rho_0)$ & $E_1$ & $E_2$
& L & $L_1$ & $L_2$ & $L_3$ & $L_4$ & $L_5$ & Ref. \\ \hline
$\mathbf{MDI}$ &  &  &  &  &  &  &  &  &  &  &  &  \\
MDI x=1 & 0.160 & -16.2 & 30.5 & 18.2 & 12.3 & 14.6 & 36.4 & -6.2 & 36.9 &
-18.6 & -33.9 & \cite{Che05} \\
MDI x=0 & 0.160 & -16.2 & 30.5 & 18.2 & 12.3 & 60.2 & 36.4 & -6.2 & 36.9 &
-18.6 & 11.6 & \cite{Che05} \\
MDI x=-1 & 0.160 & -16.2 & 30.5 & 18.2 & 12.3 & 105.7 & 36.4 & -6.2 & 36.9 &
-18.6 & 57.2 & \cite{Che05} \\
$\mathbf{Gogny}$ &  &  &  &  &  &  &  &  &  &  &  &  \\
D1 & 0.166 & -16.3 & 30.7 & 18.8 & 11.9 & 18.4 & 37.7 & -5.0 & 35.6 & -17.1
& -32.7 & \cite{Dec80} \\
D1S & 0.163 & -16.0 & 31.1 & 17.9 & 13.3 & 22.4 & 35.7 & -7.6 & 39.8 & -7.5
& -37.9 & \cite{Berger91a} \\
D1M & 0.165 & -16.0 & 28.6 & 16.8 & 11.8 & 24.8 & 33.6 & -4.2 & 35.3 & 4.5 &
-44.3 & \cite{Gor09a} \\
D1N & 0.161 & -16.0 & 29.6 & 16.5 & 13.1 & 33.6 & 33.0 & -3.7 & 39.2 & 9.2 &
-44.2 & \cite{Chap08a} \\
$\mathbf{Skyrme}$ &  &  &  &  &  &  &  &  &  &  &  &  \\
Z-fit & 0.159 & -16.0 & 26.8 & 14.5 & 12.3 & -49.7 & 29.0 & 0 & 36.9 & -13.3
& -102.4 & \cite{Fried86a} \\
Esigma-fit & 0.163 & -16.0 & 26.4 & 14.8 & 11.6 & -36.9 & 29.6 & 0 & 34.9 &
-18.1 & -83.2 & \cite{Fried86a} \\
E-fit & 0.159 & -16.1 & 27.7 & 14.1 & 13.6 & -31.3 & 28.2 & 0 & 40.7 & -15.2
& -84.9 & \cite{Fried86a} \\
Zsigma-fit & 0.163 & -15.9 & 26.7 & 15.9 & 10.8 & -29.4 & 31.8 & 0 & 32.4 &
-17.1 & -76.5 & \cite{Fried86a} \\
SVII & 0.143 & -15.8 & 27.0 & 11.4 & 15.6 & -10.2 & 22.8 & 0 & 46.7 & -11.7
& -68.0 & \cite{Giann80a} \\
SkSC4o & 0.161 & -15.9 & 27.0 & 12.3 & 14.7 & -9.7 & 24.6 & 0 & 44.0 & 0 &
-78.3 & \cite{Pea00a} \\
SVI & 0.143 & -15.8 & 26.9 & 12.0 & 14.9 & -7.3 & 24.1 & 0 & 44.6 & -12.2 &
-63.8 & \cite{Giann80a} \\
ZsigmaS-fit & 0.162 & -16.0 & 28.8 & 16.0 & 12.8 & -4.5 & 32.0 & 0 & 38.4 &
-17.9 & -57.0 & \cite{Fried86a} \\
v070 & 0.158 & -15.8 & 28.0 & 11.6 & 16.4 & -3.5 & 23.1 & 0 & 49.3 & -34.7 &
-41.2 & \cite{Pea01a} \\
v075 & 0.158 & -15.8 & 28.0 & 11.6 & 16.4 & -0.3 & 23.1 & 0 & 49.3 & -27.8 &
-45.0 & \cite{Pea01a} \\
SI & 0.155 & -16.0 & 29.2 & 13.2 & 16.0 & 1.2 & 26.4 & 0 & 48.1 & -11.4 &
-61.9 & \cite{Fried86a} \\
v080 & 0.157 & -15.8 & 28.0 & 11.6 & 16.4 & 2.2 & 23.1 & 0 & 49.3 & -21.7 &
-48.5 & \cite{Pea01a} \\
v090 & 0.157 & -15.8 & 28.0 & 11.6 & 16.4 & 5.0 & 23.1 & 0 & 49.3 & -11.6 &
-55.8 & \cite{Pea01a} \\
Skz4 & 0.160 & -16.0 & 32.0 & 17.5 & 14.5 & 5.8 & 35.1 & 0 & 43.4 & 27.9 &
-100.6 & \cite{Mar02a} \\
SkSC15 & 0.161 & -15.9 & 28.0 & 12.3 & 15.7 & 6.7 & 24.6 & 0 & 47.0 & 0 &
-65.0 & \cite{Pea00a} \\
BSk3 & 0.157 & -15.8 & 27.9 & 10.8 & 17.1 & 6.8 & 21.6 & 0 & 51.3 & -16.6 &
-49.6 & \cite{Gor03a} \\
MSk3 & 0.158 & -15.8 & 28.0 & 12.2 & 15.8 & 7.0 & 24.3 & 0 & 47.5 & 0 & -64.8
& \cite{Ton00a} \\
v105 & 0.157 & -15.8 & 28.0 & 11.6 & 16.4 & 7.1 & 23.1 & 0 & 49.3 & 0 & -65.3
& \cite{Pea01a} \\
MSk4 & 0.157 & -15.8 & 28.0 & 11.6 & 16.4 & 7.2 & 23.1 & 0 & 49.3 & 0 & -65.2
& \cite{Ton00a} \\
BSk1 & 0.157 & -15.8 & 27.8 & 11.6 & 16.2 & 7.2 & 23.1 & 0 & 48.7 & 0 & -64.7
& \cite{Sam02a} \\
v110 & 0.157 & -15.8 & 28.0 & 11.6 & 16.4 & 7.5 & 23.1 & 0 & 49.3 & 3.2 &
-68.1 & \cite{Pea01a} \\
MSk5 & 0.157 & -15.8 & 28.0 & 11.6 & 16.4 & 7.6 & 23.1 & 0 & 49.3 & 0 & -64.9
& \cite{Ton00a} \\
BSk2p & 0.157 & -15.8 & 28.0 & 11.6 & 16.4 & 7.8 & 23.2 & 0 & 49.3 & -14.9 &
-49.7 & \cite{Gor02a} \\
BSk2 & 0.157 & -15.8 & 28.0 & 11.7 & 16.3 & 8.0 & 23.3 & 0 & 49.0 & -14.8 &
-49.6 & \cite{Gor02a} \\
MSk8 & 0.158 & -15.8 & 27.9 & 11.0 & 16.9 & 8.3 & 22.1 & 0 & 50.6 & 0 & -64.5
& \cite{Gor01a} \\
v100 & 0.157 & -15.8 & 28.0 & 11.6 & 16.4 & 8.7 & 23.1 & 0 & 49.3 & -3.5 &
-60.2 & \cite{Pea01a} \\
MSk7 & 0.158 & -15.8 & 27.9 & 11.6 & 16.4 & 9.4 & 23.1 & 0 & 49.1 & 0 & -62.9
& \cite{Gor01b} \\
MSk6 & 0.157 & -15.8 & 28.0 & 11.6 & 16.4 & 9.6 & 23.1 & 0 & 49.3 & 0 & -62.8
& \cite{Ton00a} \\
SIII & 0.145 & -15.9 & 28.2 & 15.1 & 13.1 & 9.9 & 30.2 & 0 & 39.2 & -14.8 &
-44.7 & \cite{Bei75a} \\
MSk9 & 0.158 & -15.8 & 28.0 & 12.2 & 15.8 & 10.4 & 24.3 & 0 & 47.5 & 0.0 &
-61.5 & \cite{Gor01a} \\
BSk4 & 0.157 & -15.8 & 28.0 & 13.2 & 14.8 & 12.5 & 26.4 & 0 & 44.4 & -6.5 &
-51.7 & \cite{Gor03a} \\
Skz3 & 0.160 & -16.0 & 32.0 & 17.5 & 14.5 & 13.0 & 35.1 & 0 & 43.4 & 10.3 &
-75.8 & \cite{Mar02a} \\
BSk8 & 0.159 & -15.8 & 28.0 & 15.3 & 12.7 & 14.9 & 30.6 & 0 & 38.2 & 7.5 &
-61.4 & \cite{Sam04a} \\
Dutta & 0.162 & -16.0 & 26.6 & 12.4 & 14.2 & 16.5 & 24.8 & 0 & 42.7 & 0 &
-51.0 & \cite{Dut86a} \\
Skz2 & 0.160 & -16.0 & 32.0 & 17.5 & 14.5 & 16.8 & 35.1 & 0 & 43.4 & -10.7 &
-50.9 & \cite{Mar02a} \\
BSk6 & 0.157 & -15.8 & 28.0 & 15.2 & 12.8 & 16.8 & 30.4 & 0 & 38.4 & 6.2 &
-58.2 & \cite{Gor03a} \\
BSk7 & 0.157 & -15.8 & 28.0 & 15.2 & 12.8 & 18.0 & 30.4 & 0 & 38.4 & 7.5 &
-58.3 & \cite{Gor03a} \\
SKP & 0.163 & -16.0 & 30.0 & 12.4 & 17.6 & 19.6 & 24.8 & 0 & 52.7 & -26.1 &
-31.9 & \cite{bro98a} \\
BSk5 & 0.157 & -15.8 & 28.7 & 13.2 & 15.5 & 21.4 & 26.4 & 0 & 46.5 & -8.0 &
-43.5 & \cite{Gor03a} \\
Skz1 & 0.160 & -16.0 & 32.0 & 17.5 & 14.5 & 27.7 & 35.1 & 0 & 43.4 & -24.7 &
-26.1 & \cite{Mar02a} \\ \hline\hline
\end{tabular*}%
\end{table*}

\begin{table*}[tbp]
\renewcommand{\tablename}{} TABLE I (Continued.)\newline
\begin{tabular*}{\textwidth}{@{\extracolsep{\fill}}lcccccccccccr}
\toprule Model & $\rho_0$ & $E_0(\rho_0)$ & $E_{sym}(\rho_0)$ & $E_1$ & $E_2$
& L & $L_1$ & $L_2$ & $L_3$ & $L_4$ & $L_5$ & Ref. \\ \hline
SIIIs & 0.148 & -16.1 & 32.0 & 15.0 & 17.0 & 28.7 & 29.9 & 0 & 51.0 & -5.0 &
-47.3 & \cite{Giann80a} \\
SKT6 & 0.161 & -16.0 & 30.0 & 12.3 & 17.6 & 30.9 & 24.7 & 0 & 52.9 & 0 &
-46.7 & \cite{Ton84a} \\
SKT7 & 0.161 & -15.9 & 29.5 & 14.8 & 14.7 & 31.1 & 29.6 & 0 & 44.2 & -14.8 &
-27.8 & \cite{Ton84a} \\
SKXm & 0.159 & -16.0 & 31.2 & 12.7 & 18.5 & 32.1 & 25.3 & 0 & 55.6 & -22.0 &
-26.9 & \cite{bro98a} \\
RATP & 0.160 & -16.0 & 29.2 & 18.4 & 10.9 & 32.4 & 36.8 & 0 & 32.6 & -20.5 &
-16.5 & \cite{Rayet} \\
SkSC14 & 0.161 & -15.9 & 30.0 & 12.3 & 17.7 & 33.1 & 24.6 & 0 & 53.0 & 0 &
-44.5 & \cite{Pea00a} \\
SKX & 0.155 & -16.1 & 31.1 & 12.1 & 19.0 & 33.2 & 24.3 & 0 & 56.9 & -23.7 &
-24.3 & \cite{bro98a} \\
MSk2 & 0.157 & -15.8 & 30.0 & 11.6 & 18.4 & 33.4 & 23.1 & 0 & 55.3 & 0.0 &
-45.1 & \cite{Ton00a} \\
SKXce & 0.155 & -15.9 & 30.1 & 12.0 & 18.2 & 33.5 & 23.9 & 0 & 54.5 & -24.2
& -20.8 & \cite{bro98a} \\
BSk15 & 0.159 & -16.0 & 30.0 & 15.3 & 14.7 & 33.6 & 30.6 & 0.0 &
44.2 & -3.9 & -37.2 & \cite{Gor08a} \\
SKT8 & 0.161 & -15.9 & 29.9 & 14.8 & 15.1 & 33.7 & 29.6 & 0 & 45.4 &
0 & -41.3 & \cite{Ton84a} \\
SKT9 & 0.160 & -15.9 & 29.8 & 14.8 & 15.0 & 33.7 & 29.5 & 0 & 45.0 & 0 &
-40.8 & \cite{Ton84a} \\
MSk1 & 0.157 & -15.8 & 30.0 & 12.2 & 17.8 & 33.9 & 24.3 & 0 & 53.5 & 0 &
-43.9 & \cite{Ton00a} \\
BSk16 & 0.159 & -16.1 & 30.0 & 15.3 & 14.7 & 34.9 & 30.5 & 0.0 &
44.2 & -2.1 & -37.8 & \cite{Cha08a} \\
Skz0 & 0.160 & -16.0 & 32.0 & 17.5 & 14.5 & 35.1 & 35.1 & 0 & 43.4 &
-42.1 & -1.3 & \cite{Mar02a} \\
Skyrme1p & 0.155 & -16.0 & 29.4 & 13.2 & 16.1 & 35.3 & 26.4 & 0 & 48.4 &
-11.4 & -28.1 & \cite{Pet95a} \\
BSk17 & 0.159 & -16.1 & 30.0 & 15.3 & 14.7 & 36.3 & 30.5 & 0.0 &
44.2 & -2.1 & -36.4 & \cite{Gor09b} \\
BSk10 & 0.159 & -15.9 & 30.0 & 13.3 & 16.7 & 37.2 & 26.6 & 0 & 50.1
& -10.5 & -29.0 & \cite{Gor06a} \\
SGII & 0.158 & -15.6 & 26.8 & 15.5 & 11.3 & 37.6 & 31.0 & 0 & 33.9 & -16.1 &
-11.2 & \cite{Giai81a} \\
BSk12 & 0.159 & -15.9 & 30.0 & 13.3 & 16.7 & 38.0 & 26.5 & 0 & 50.2 & -9.9 &
-28.8 & \cite{Gor06a} \\
BSk11 & 0.159 & -15.9 & 30.0 & 13.3 & 16.7 & 38.4 & 26.5 & 0 & 50.2 & -9.8 &
-28.6 & \cite{Gor06a} \\
SLy10 & 0.156 & -15.9 & 32.0 & 17.6 & 14.3 & 38.7 & 35.3 & 0 & 43.0 & 15.4 &
-55.0 & \cite{Chaba95a} \\
BSk13 & 0.159 & -15.9 & 30.0 & 13.3 & 16.7 & 38.8 & 26.5 & 0 & 50.2 & -9.7 &
-28.2 & \cite{Gor06a} \\
BSk9 & 0.159 & -15.9 & 30.0 & 15.3 & 14.7 & 39.9 & 30.6 & 0 & 44.2 &
10.8 & -45.7 & \cite{Sam04a} \\
KDE & 0.164 & -16.0 & 32.0 & 16.5 & 15.4 & 41.4 & 33.1 & 0 & 46.3 & 12.3 &
-50.2 & \cite{Agra05a} \\
BSk14 & 0.159 & -15.9 & 30.0 & 15.3 & 14.7 & 43.9 & 30.5 & 0.0 &
44.2 & -2.0 & -28.8 & \cite{Gor07a} \\
SLy230a & 0.160 & -16.0 & 32.0 & 17.6 & 14.4 & 44.3 & 35.2 & 0 &
43.1 & 32.0 & -66.1 & \cite{Chaba98a} \\
KDE0 & 0.161 & -16.1 & 33.0 & 17.2 & 15.8 & 45.2 & 34.4 & 0 & 47.4 & 6.9 &
-43.5 & \cite{Agra05a} \\
SLy8 & 0.160 & -16.0 & 31.4 & 17.7 & 13.8 & 45.3 & 35.3 & 0 & 41.3 &
13.7 & -45.1 & \cite{Chaba95a} \\
SLy4 & 0.160 & -16.0 & 31.8 & 17.6 & 14.2 & 45.4 & 35.3 & 0 & 42.5 & 13.8 &
-46.2 & \cite{Chaba98a} \\
SLy0 & 0.161 & -16.0 & 31.5 & 17.6 & 13.8 & 45.4 & 35.3 & 0 & 41.5 & 13.6 &
-45.0 & \cite{Chaba95a} \\
SLy3 & 0.160 & -16.0 & 32.1 & 17.7 & 14.4 & 45.5 & 35.3 & 0 & 43.2 & 13.8 &
-46.8 & \cite{Chaba95a} \\
SKMs & 0.160 & -15.8 & 30.0 & 15.6 & 14.4 & 45.8 & 31.2 & 0 & 43.3 & -19.4 &
-9.3 & \cite{Bart82a} \\
SLy230b & 0.160 & -16.0 & 32.0 & 17.6 & 14.4 & 46.0 & 35.3 & 0 & 43.1 & 13.9
& -46.4 & \cite{Chaba97a} \\
SLy7 & 0.158 & -15.9 & 32.0 & 17.7 & 14.3 & 47.2 & 35.4 & 0 & 42.8 & 15.0 &
-46.0 & \cite{Chaba98a} \\
SLy6 & 0.159 & -15.9 & 32.0 & 17.7 & 14.2 & 47.4 & 35.4 & 0 & 42.7 & 14.6 &
-45.3 & \cite{Chaba98a} \\
SKb & 0.155 & -16.0 & 23.9 & 19.8 & 4.1 & 47.5 & 39.6 & 0 & 12.3 & -21.6 &
17.3 & \cite{Giai81a} \\
SLy5 & 0.160 & -16.0 & 32.0 & 17.6 & 14.4 & 48.3 & 35.3 & 0 & 43.1 & 13.7 &
-43.8 & \cite{Chaba98a} \\
SLy2 & 0.160 & -15.9 & 32.3 & 17.6 & 14.7 & 48.8 & 35.2 & 0 & 44.0 & 13.6 &
-44.1 & \cite{Chaba95a} \\
SLy1 & 0.160 & -16.0 & 32.5 & 17.6 & 14.9 & 48.8 & 35.2 & 0 & 44.7 & 13.7 &
-44.8 & \cite{Chaba95a} \\
SKM & 0.160 & -15.8 & 30.7 & 15.6 & 15.2 & 49.3 & 31.2 & 0 & 45.5 & -18.0 &
-9.3 & \cite{Kri80a} \\
SII & 0.148 & -16.0 & 34.2 & 20.1 & 14.0 & 50.0 & 40.3 & 0 & 42.0 & -19.2 &
-13.1 & \cite{BV1972} \\
Skzm1 & 0.160 & -16.0 & 32.0 & 17.5 & 14.5 & 54.1 & 35.1 & 0 & 43.4 & -47.9
& 23.6 & \cite{Mar02a} \\
SKT3 & 0.161 & -15.9 & 31.5 & 12.3 & 19.2 & 55.3 & 24.7 & 0 & 57.5 & 0 &
-26.8 & \cite{Ton84a} \\
SLy9 & 0.151 & -15.8 & 32.1 & 17.8 & 14.4 & 55.4 & 35.5 & 0 & 43.2 & 17.9 &
-41.1 & \cite{Chaba95a} \\
SKT3s & 0.160 & -16.0 & 31.7 & 12.3 & 19.4 & 55.9 & 24.6 & 0 & 58.2 & 0 &
-26.9 & \cite{Ton84a} \\
SKT1s & 0.160 & -16.0 & 32.0 & 12.3 & 19.7 & 56.1 & 24.6 & 0 & 59.2 & 0 &
-27.7 & \cite{Ton84a} \\
SKT2 & 0.161 & -15.9 & 32.0 & 12.3 & 19.7 & 56.2 & 24.7 & 0 & 59.0 & 0 &
-27.5 & \cite{Ton84a} \\
SKT1 & 0.161 & -16.0 & 32.0 & 12.3 & 19.7 & 56.2 & 24.7 & 0 & 59.1 & 0 &
-27.5 & \cite{Ton84a} \\
MSkA & 0.153 & -16.0 & 30.3 & 15.1 & 15.3 & 57.2 & 30.1 & 0 & 45.9 & -16.4 &
-2.4 & \cite{Shar95a} \\
SkI6 & 0.159 & -15.9 & 29.9 & 19.1 & 10.8 & 59.2 & 38.2 & 0 & 32.4 & 23.1 &
-34.4 & \cite{Naza96a} \\
MSL0 & 0.160 & -16.0 & 30.0 & 15.4 & 14.6 & 60.0 & 30.7 & 0 & 43.9 & -13.2 &
-1.5 & \cite{Che10} \\
SkI4 & 0.160 & -15.9 & 29.5 & 18.9 & 10.6 & 60.4 & 37.8 & 0 & 31.7 & 21.4 &
-30.6 & \cite{Rein95a} \\
LNS & 0.175 & -15.3 & 33.4 & 15.8 & 17.7 & 61.5 & 31.5 & 0 & 53.0 & -12.8 &
-10.2 & \cite{Cao06a} \\ \hline\hline
\end{tabular*}%
\end{table*}

\begin{table*}[tbp]
\renewcommand{\tablename}{} TABLE I (Continued.)\newline
\begin{tabular*}{\textwidth}{@{\extracolsep{\fill}}lcccccccccccr}
\toprule Model & $\rho_0$ & $E_0(\rho_0)$ & $E_{sym}(\rho_0)$ & $E_1$ & $E_2$
& L & $L_1$ & $L_2$ & $L_3$ & $L_4$ & $L_5$ & Ref. \\ \hline
SIV & 0.151 & -16.0 & 31.2 & 25.1 & 6.1 & 63.5 & 50.2 & 0 & 18.4 & -23.5 &
18.5 & \cite{Bei75a} \\
SGI & 0.154 & -15.9 & 28.3 & 19.7 & 8.6 & 63.9 & 39.5 & 0 & 25.8 & -7.0 & 5.6
& \cite{Giai81a} \\
SKOs & 0.160 & -15.8 & 31.9 & 13.7 & 18.2 & 68.9 & 27.4 & 0 & 54.7 & -2.4 &
-10.8 & \cite{Rein99a} \\
SkMP & 0.157 & -15.6 & 29.9 & 18.6 & 11.3 & 70.3 & 37.1 & 0 & 34.0 & -13.1 &
12.3 & \cite{Benn89a} \\
Ska & 0.155 & -16.0 & 32.9 & 19.8 & 13.1 & 74.6 & 39.6 & 0 & 39.4 & -21.6 &
17.3 & \cite{Koh76a} \\
SKO & 0.160 & -15.8 & 32.0 & 13.7 & 18.2 & 79.1 & 27.5 & 0 & 54.7 & -4.3 &
1.3 & \cite{Rein99a} \\
SKYT & 0.148 & -15.4 & 33.7 & 19.3 & 14.3 & 80.8 & 38.7 & 0 & 43.0 & -10.8 &
9.9 & \cite{Ko74a} \\
Rsigma-fit & 0.158 & -15.6 & 30.6 & 15.5 & 15.0 & 85.7 & 31.1 & 0.0
& 45.1 & -14.6 & 24.1 & \cite{Fried86a} \\
SK272 & 0.155 & -16.3 & 37.4 & 15.6 & 21.8 & 91.7 & 31.2 & 0 & 65.4
& -17.6 & 12.6 & \cite{Agra03a} \\
Gsigma-fit & 0.158 & -15.6 & 31.4 & 15.5 & 15.9 & 94.0 & 31.0 & 0.0
& 47.6 & -14.6 & 30.0 & \cite{Fried86a} \\
SKT4 & 0.159 & -16.0 & 35.5 & 12.2 & 23.2 & 94.1 & 24.5 & 0 & 69.7 &
0 & 0 & \cite{Ton84a} \\
SK255 & 0.157 & -16.3 & 37.4 & 15.2 & 22.2 & 95.1 & 30.5 & 0 & 66.5 & -20.9
& 19.0 & \cite{Agra03a} \\
SV & 0.155 & -16.0 & 32.8 & 31.4 & 1.4 & 96.1 & 62.8 & 0 & 4.2 & -29.1 & 58.2
& \cite{Fried86a} \\
SKT5 & 0.164 & -16.0 & 37.0 & 12.5 & 24.5 & 98.5 & 25.0 & 0 & 73.6 & 0 & 0 &
\cite{Ton84a} \\
SkI3 & 0.158 & -16.0 & 34.8 & 21.1 & 13.8 & 100.5 & 42.1 & 0 & 41.3 & 35.5 &
-18.4 & \cite{Rein95a} \\
SkI2 & 0.158 & -15.8 & 33.4 & 17.7 & 15.6 & 104.3 & 35.5 & 0 & 46.9 & 15.7 &
6.2 & \cite{Rein95a} \\
SkI5 & 0.156 & -15.8 & 36.6 & 20.8 & 15.8 & 129.3 & 41.7 & 0 & 47.4 & 35.0 &
5.2 & \cite{Rein95a} \\
SkI1 & 0.160 & -16.0 & 37.5 & 17.7 & 19.8 & 161.1 & 35.5 & 0 & 59.3 & 14.2 &
52.0 & \cite{Rein95a} \\ \hline\hline
\end{tabular*}%
\end{table*}

Due to the special interest on the values of $E_{sym}(\rho )$ and $L(\rho )$
at $\rho _{0}$, we list in Table \ref{TableI} the values of the
characteristic parameters $\rho _{0}$, $E_{0}(\rho _{0})$, $E_{sym}(\rho
_{0})$, $E_{1}$, $E_{2}$, $L$, $L_{1}$, $L_{2}$, $L_{3}$, $L_{4}$ and $L_{5}$
at $\rho _{0}$ for the MDI interaction with $x=-1$, $0$ and $1$, the
Gogny-Hartree-Fock predictions with D1, D1S, D1N, and D1M as well as the
Skyrme-Hartree-Fock predictions with $112$ standard Skyrme interactions (In
the table, the interactions in different models are in order according to
the $L$ value and the corresponding reference for different interactions is
included in the last column). For the MDI interaction model, one can see
from Table \ref{TableI} that the value of $E_{2}(\rho _{0})$ is comparable
with that of $E_{1}(\rho _{0})$, and all the $E_{0}(\rho _{0})$, $E_{1}(\rho
_{0})$, and $E_{2}(\rho _{0})$ are the same for different $x$ values due to
the fact that the $U_{0}(\rho ,k)$ and $U_{sym,1}(\rho _{0},k)$ are
independent of the $x$ parameter by construction \cite{Che05}. Because of
the same reason, the values of $L_{1}$, $L_{2}$, $L_{3}$, $L_{4}$ are all
independent of the $x$ parameter too. Therefore, for the MDI interaction,
the $x$ dependence of the $L$ parameter is completely determined by the
second-order symmetry potential $U_{sym,2}(\rho ,k_{F})$. Depending on the
value of the $x$ parameter, the $U_{sym,2}(\rho ,k_{F})$ can be positive or
negative. In particular, we have $L=L_{1}+L_{2}+L_{3}+L_{4}=48.5$ MeV if we
assume $U_{sym,2}(\rho ,k_{F})=0$. Furthermore, it is seen that the $L_{2}$
contribution is relatively small compared with that of $L_{1}$, $L_{3}$, or $%
L_{4}$, indicating that the contribution due to the momentum dependence of
the nucleon effective mass is unimportant, and this is consistent with the
observation from Fig. \ref{LMDI}.

For the Gogny interaction, the value of $L$ listed in Table \ref{TableI}
ranges from about $18$ MeV to $34$ MeV. And similarly with the MDI\
interaction model, the $L_{2}$ is relatively small, ranging from about $-8$
MeV to $-3$ MeV. It is interesting to see that the values of $L_{1}$ and $%
L_{3}$ from the Gogny interactions are quite similar with those of the MDI\
interaction model, and thus the difference of the $L$ parameter from
different interactions in these two models is mainly due to the variation of
the $L_{4}$ and $L_{5}$. Furthermore, for different Gogny interactions
considered here, the value of $L_{4}$ can be positive or negative while the
value of $L_{5}$ is always negative, and the $L_{5}$ contribution to the $L$
parameter usually is relatively important.

For the standard Skyrme interactions, we have $L_{2}=0$ MeV. For the $112$
Skyrme interactions considered in Table \ref{TableI}, it is seen that the
value of $L$ ranges from about $-50$ MeV to $160$ MeV, $L_{1}$ from about $%
20 $ MeV to $60$ MeV, $L_{3}$ from about $4$ MeV to $74$ MeV, $L_{4}$ from
about $-48$ MeV to $36$ MeV, and $L_{5}$ from about $-102$ MeV to $58$ MeV.
Therefore, the different term contributions to the $L$ parameter can change
a lot in the standard Skyrme interactions, especially for $L_{3}$, $L_{4}$,
and $L_{5}$.
\begin{figure}[tbh]
\includegraphics[width=8.5cm]{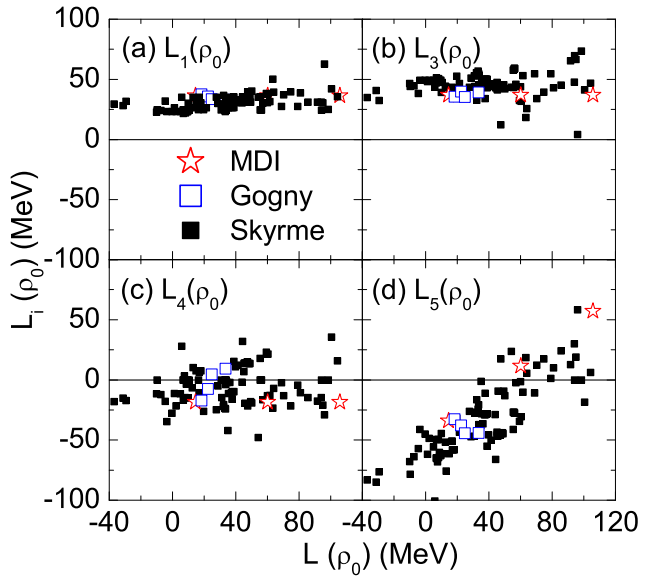}
\caption{(Color online) Correlations of $L_{1}(\protect\rho _{0})$ (a), $%
L_{3}(\protect\rho _{0})$ (b), $L_{4}(\protect\rho _{0})$ (c), and $L_{5}(%
\protect\rho _{0})$ (d) with $L(\protect\rho _{0})$ for the MDI interaction
with $x=-1$, $0$ and $1$, Gogny-Hartree-Fock predictions with D1, D1S, D1N
and D1M as well as Skyrme-Hartree-Fock predictions with the $112$ standard
Skyrme interactions considered in Table \protect\ref{TableI}.}
\label{L15SHF}
\end{figure}

In order to see more clearly and intuitively the different term
contributions to the $L$ parameter, we show in Fig. \ref{L15SHF} the
correlations of $L_{1}(\rho _{0})$, $L_{3}(\rho _{0})$, $L_{4}(\rho _{0})$,
and $L_{5}(\rho _{0})$ with $L(\rho _{0})$ for the MDI interaction with $%
x=-1 $, $0$ and $1$, the Gogny-Hartree-Fock predictions with D1, D1S, D1N,
and D1M as well as the Skyrme-Hartree-Fock predictions with the $112$
standard Skyrme interactions. One can see from Fig. \ref{L15SHF} that the
results from the MDI interaction model and the Gogny-Hartree-Fock
calculations are essentially consistent with the systematics of the
Skyrme-Hartree-Fock predictions. Based on these calculated results, we find
that a statistical analysis can lead to $L_{1}(\rho _{0})\approx 30\pm 6.5$
MeV, $L_{3}(\rho _{0})\approx 46\pm 9.5$ MeV, $L_{4}(\rho _{0})\approx -4\pm
15$ MeV, and $L_{5}(\rho _{0})\approx -35\pm 30$ MeV. These results indicate
that, within the standard Skyrme-Hartree-Fock energy density functional, $%
L_{1}(\rho _{0}) $ and $L_{3}(\rho _{0})$ are relatively well constrained,
and the main uncertainties are due to the $L_{4}(\rho _{0})$ and $L_{5}(\rho
_{0})$ contributions. Furthermore, it is interesting to see from Fig. \ref%
{L15SHF} that there exhibits an approximately linear correlation between $%
L_{5}(\rho _{0})$ and $L(\rho _{0})$. If we use the present empirical
constraint $L(\rho _{0})=60\pm 30$ MeV, then we find the $L_{5}(\rho _{0})$
can vary from about $-66$ MeV to $24$ MeV, i.e., the value of $%
U_{sym,2}(\rho _{0},k_{F})$ can vary from about $-22$ MeV to $8$ MeV.

\subsection{The symmetry potential $U_{sym,1}(\protect\rho ,k)$}

\begin{figure}[tbh]
\includegraphics[width=8.6cm]{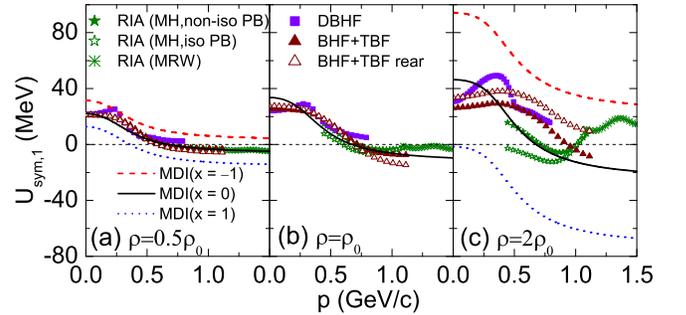}
\caption{(Color online) Momentum dependence of the $U_{sym,1}(\protect\rho %
,k)$ at $\protect\rho =0.5\protect\rho _{0}$ (a), $\protect\rho _{0}$ (b)
and $2\protect\rho _{0}$ (c) using the MDI interaction with $x=-1$, $0$, and
$1$. The corresponding results from several microscopic approaches are also
included for comparison (See the text for details).}
\label{Usym1MDI}
\end{figure}

Shown in Fig. \ref{Usym1MDI} is the momentum dependence of the $%
U_{sym,1}(\rho ,k)$ at $\rho =0.5\rho _{0}$, $\rho _{0}$ and $2\rho
_{0}$ using the MDI interaction with $x=-1$, $0$, and $1$. For
comparison, we also include in Fig. \ref{Usym1MDI} the corresponding
results from several microscopic approaches, including the
relativistic impulse approximation (RIA) \cite{Che05c,LiZH06} using
the empirical nucleon-nucleon scattering amplitude determined by
Murdock and Horowitz (MH) \cite{Mur87} with isospin-dependent and
isospin-independent Pauli blocking corrections as well as by McNeil,
Ray, and Wallace (MRW) \cite{McN83}, the relativistic
Dirac-Brueckner-Hartree-Fock (DBHF) theory \cite{Dal05}, and the
non-relativistic Brueckner-Hartree-Fock (BHF) theory with and
without the three-body force (TBF) rearrangement contribution
\cite{Zuo06}. For these microscopic results,
one can see that they are all consistent with each other around and below $%
\rho _{0}$ although there still exist larger uncertainties at higher density
of $\rho =2\rho _{0}$. It is interesting to see that the momentum dependence
of the $U_{sym,1}(\rho ,k) $ from the MDI interaction with $x=0$ are in good
agreement with the results from the microscopic approaches. It should be
noted that the momentum dependence of the $U_{sym,1}(\rho ,k)$ at $\rho _{0}$
is the same for $x=-1$, $0$, and $1$ since $U_{sym,1}(\rho ,k)$ is
independent of the $x$ parameter at $\rho _{0}$ by construction as mentioned
previously.
\begin{figure}[tbh]
\includegraphics[width=8.6cm]{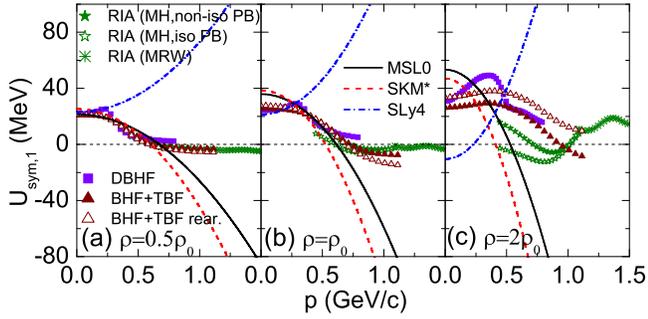}
\caption{(Color online) Same as Fig. \protect\ref{Usym1MDI} but for the
Skyrme Hartree-Fock\ approach with MSL0, SLy4, and SKM*.}
\label{Usym1SHF}
\end{figure}

\begin{figure}[tbh]
\includegraphics[width=8.6cm]{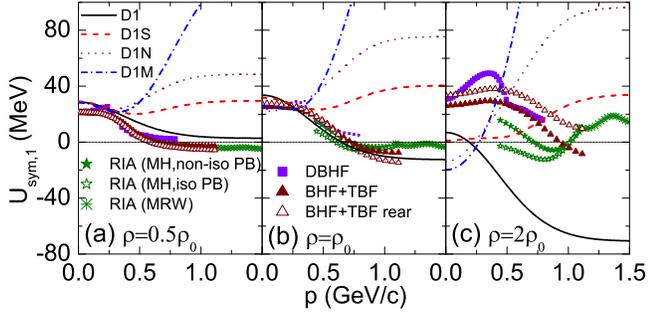}
\caption{(Color online) Same as Fig. \protect\ref{Usym1MDI} but for Gogny
Hartree-Fock\ approach with D1, D1S, D1N, and D1M.}
\label{Usym1Gogny}
\end{figure}

Similarly as in Fig. \ref{Usym1MDI}, we plot in Fig. \ref{Usym1SHF} and Fig. %
\ref{Usym1Gogny} the the momentum dependence of the $U_{sym,1}(\rho ,k)$ at $%
\rho =0.5\rho _{0}$, $\rho _{0}$ and $2\rho _{0}$ using the Skyrme
Hartree-Fock\ approach with MSL0, SLy4 and SKM*, and the Gogny Hartree-Fock\
approach with D1, D1S, D1N and D1M, respectively. On the one hand, it is
seen from Fig. \ref{Usym1SHF} that, for $0<p<700$ MeV/c, the results from
the MSL0 and SKM* interactions agree well with those from microscopic
approaches while the results from the SLy4 interaction seem to display large
deviation from the microscopic results, especially at higher nucleon
momenta. On the other hand, one can see from Fig. \ref{Usym1Gogny} that at $%
\rho =0.5\rho _{0}$ and $\rho _{0}$, the results from D1 interaction are
consistent with those of the microscopic calculations while the results from
D1S, D1N and D1M exhibit large deviation from the the microscopic
calculations except at low nucleon momenta ($p\lesssim 300$ MeV/c). At $\rho
=2\rho _{0}$, the Gogny Hartree-Fock\ calculations display different results
from the microscopic ones and the strong model dependence appears.

Overall, one can see from Fig. \ref{Usym1MDI}, Fig. \ref{Usym1SHF} and Fig. %
\ref{Usym1Gogny} that the momentum dependence of the $U_{sym,1}$ varies from
one interaction to another for the MDI interaction model, the Skyrme
Hartree-Fock\ approach and the Gogny Hartree-Fock\ approach. For the MDI
interaction with $x=-1$, $0$, and $1$, the Gogny interaction with D1, and
the Skyrme interaction with MSL0 and SKM*, the value of $U_{sym,1}$ can be
negative at higher momentum while for the Gogny interaction with D1S, D1N,
D1M and the Skyrme interaction with SLy4, the value of $U_{sym,1}$ is
positive at higher momentum.

\subsection{The second-order symmetry potential $U_{sym,2}(\protect\rho ,k)$}

\begin{figure}[tbh]
\includegraphics[width=8.6cm]{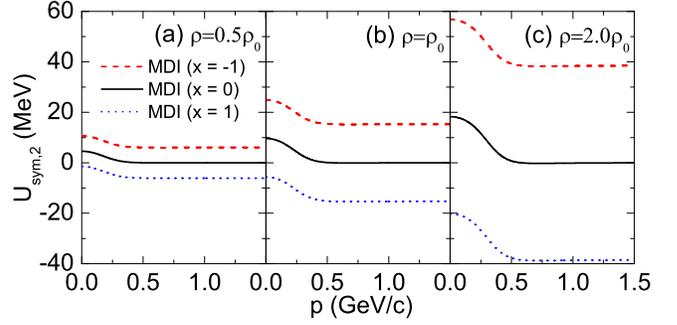}
\caption{(Color online) Momentum dependence of the $U_{sym,2}(\protect\rho %
,k)$ at $\protect\rho =0.5\protect\rho _{0}$ (a), $\protect\rho _{0}$ (b)
and $2\protect\rho _{0}$ (c) using the MDI interaction with $x=-1$, $0$, and
$1$.}
\label{Usym2MDI}
\end{figure}

\begin{figure}[tbh]
\includegraphics[width=8.6cm]{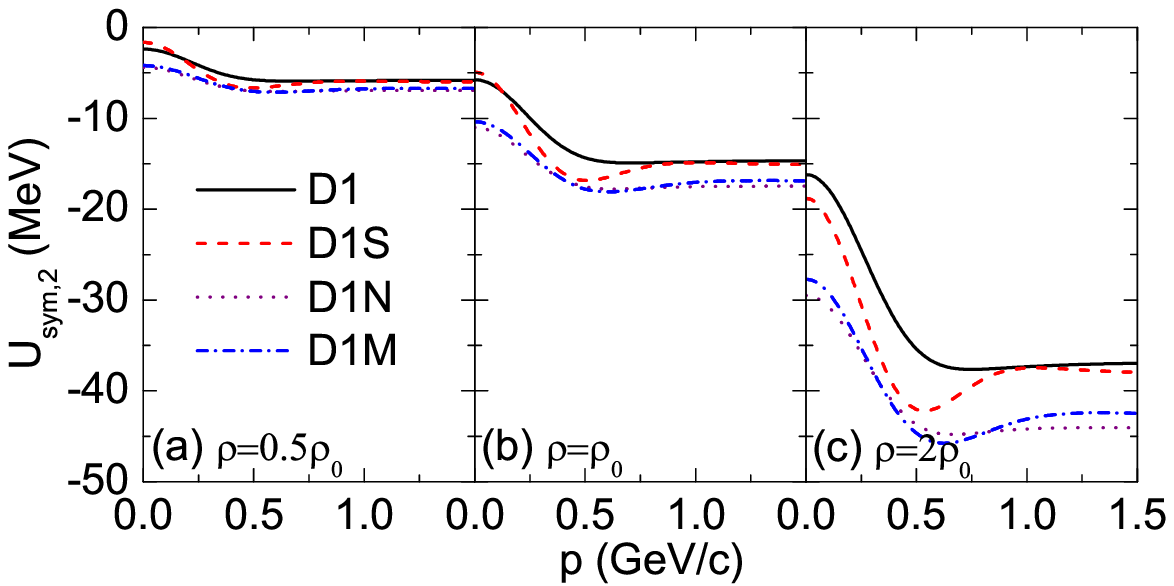}
\caption{(Color online) Same as Fig. \protect\ref{Usym2MDI} but for Gogny
Hartree-Fock\ approach with D1, D1S, D1N, and D1M.}
\label{Usym2Gogny}
\end{figure}

\begin{figure}[tbh]
\includegraphics[width=8.3cm]{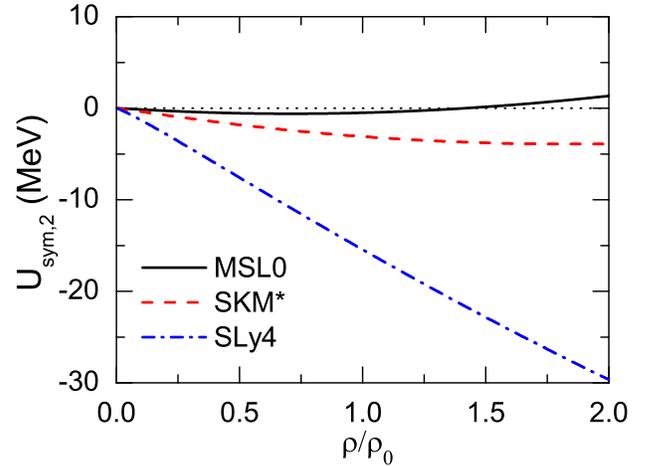}
\caption{(Color online) Density dependence of $U_{sym,2}$ for the Skyrme
Hartree-Fock\ approach with MSL0, SLy4, and SKM*.}
\label{Usym2SHF}
\end{figure}

As for $U_{sym,2}$, to our best knowledge, there is not any
experimental/empirical information or theoretical predictions so far. Fig.~%
\ref{Usym2MDI} and Fig.~\ref{Usym2Gogny} show the momentum dependence of $%
U_{sym,2}$ at $0.5\rho _{0}$, $\rho _{0}$ and $2\rho _{0}$ in the MDI
interaction model and the Gogny Hartree-Fock approach, respectively. Since $%
U_{sym,2}$ is independent of the nucleon momentum in the Skyrme Hartree-Fock
approach, we only show here its density dependence in Fig.~\ref{Usym2SHF}.
From Fig.~\ref{Usym2MDI} and Fig.~\ref{Usym2Gogny}, it is interesting to see
that for all interactions in the MDI interaction model and the Gogny
Hartree-Fock approach at $\rho =0.5\rho _{0}$, $\rho _{0}$ and $2\rho _{0}$,
$U_{sym,2}$ firstly decreases with nucleon momentum and then essentially
saturates when the nucleon momentum is larger than about ${500}$ MeV/c.
Especially, the results from the MDI interaction with ${x=1}$ seem to be in
quantitative agreement with those from the Gogny Hartree-Fock approach.
Furthermore, one can see from Fig.~\ref{Usym2MDI} and Fig.~\ref{Usym2Gogny}
that the magnitude of $U_{sym,2}$ increases with the density, and this is
also true for the Skyrme Hartree-Fock approach as shown in Fig.~\ref%
{Usym2SHF}. Another interesting feature is that for the MDI interaction
model and the Skyrme Hartree-Fock approach, $U_{sym,2}$ can be either
negative or positive while it is always negative for the Gogny Hartree-Fock
approach with the interactions considered here. Therefore, any experimental
constraints about $U_{sym,2}$ will be very useful and important for
constraining the theoretical models.


Comparing Fig.~\ref{Usym1MDI} with Fig.~\ref{Usym2MDI} for the MDI
interaction model, Fig.~\ref{Usym1SHF} with Fig.~\ref{Usym2SHF} for the
Skyrme Hartree-Fock approach, and Fig.~\ref{Usym1Gogny} with Fig.~\ref%
{Usym2Gogny} for the Gogny Hartree-Fock approach, one can see that, at fixed
density and momentum, the magnitude of $U_{sym,2}$ is generally comparable
with that of $U_{sym,1}$. So, in Eq.~(\ref{UtauTaylor}), compared with $%
U_{sym,1}(\rho ,k)\delta $, the contribution from $U_{sym,2}(\rho ,k)\delta
^{2}$ could be negligible only if $\delta $ is small ($\delta \ll 1$).
Therefore, we conclude from the present model calculations that, when $%
\delta $ is small ($\delta \ll 1$), the Lane potential might be a good
approximation to the single-nucleon potential $U_{n/p}(\rho ,\delta ,p)$.
However, the contributions of $U_{sym,2}(\rho ,k)$ might not be simply
neglected when $\delta $ is close to $1$.

\section{Summary}

\label{summary}

Using the Hugenholtz--Van Hove theorem, we have explicitly and analytically
expressed the symmetry energy $E_{sym}(\rho )$ and its density slope $L(\rho
)$ in terms of the single-nucleon potential in asymmetric nuclear matter
that might be extracted from experiments. We have carefully checked the
contributions of each decomposed term, i.e., $E_{1}(\rho )$ and $E_{2}(\rho
) $ for $E_{sym}(\rho )$, and $L_{1}(\rho )$, $L_{2}(\rho )$, $L_{3}(\rho )$%
, $L_{4}(\rho )$ and $L_{5}(\rho )$ for $L(\rho )$ by using three popular
phenomenological nuclear interaction models in nuclear structure and
reaction studies, namely, the isospin and momentum dependent MDI model, the
Skyrme Hartree-Fock approach, and the Gogny Hartree-Fock approach.

Our results have indicated that the $E_{2}(\rho )$ due to the symmetry
potential $U_{sym,1}(\rho ,k_{F})$ is comparable with the $E_{1}(\rho )$
which describes the kinetic part including the nucleon effective mass
contribution and the observed different density behaviors of $E_{sym}(\rho )$
for different interactions are essentially due to the variation of the
symmetry potential $U_{sym,1}(\rho ,k)$. For the density slope parameter $%
L(\rho )$, interestingly, we have found that, although the term $L_{2}$ due
to the momentum dependence of the nucleon effective mass might not have
significant contributions. the term $L_{5}$, which is from the second-order
symmetry potential $U_{sym,2}(\rho ,k_{F})$, generally cannot be negligible.

By analyzing the density and momentum dependence of $U_{sym,1}(\rho ,k)$ and
$U_{sym,2}(\rho ,k)$ for the three nuclear effective interaction models, we
have demonstrated that the magnitude of the second-order symmetry potential $%
U_{sym,2}(\rho ,k)$ is generally comparable with that of the symmetry
potential $U_{sym,1}(\rho ,k)$ and thus the Lane potential $U_{n/p}(\rho
,\delta ,k)\approx U_{0}(\rho ,k)\pm U_{sym,1}(\rho ,k)\delta $ could be a
good approximation to the single-nucleon potential only if the isospin
asymmetry $\delta $ is small ($\delta \ll 1$). However, $U_{sym,2}(\rho ,k)$
might not be neglected in describing the single-nucleon potential $%
U_{n/p}(\rho ,\delta ,k)$ in extremely neutron(proton)-rich nuclear matter,
e.g., in neutron stars and the neutron-skin region of heavy nuclei, where
the value of $\delta $ could be very large (close to $1$).

While the momentum dependence of $U_{0}(\rho ,k)$ and
$U_{sym,1}(\rho ,k)$ has been extensively investigated and
relatively well constrained from the measured nucleon optical model
potentials, heavy ion collisions experiments, and the microscopic
calculations, especially around and below nuclear matter saturation
density, knowledge on $U_{sym,2}(\rho ,k)$ is still very poorly
known. The results on $U_{sym,2}(\rho ,k)$ presented here from the
three phenomenological models have indicated large model dependence.
Therefore, to constrain $U_{sym,2}(\rho ,k)$ from experiments or
microscopic calculations (e.g., BHF and DBHF) based on
nucleon-nucleon interactions derived from scattering phase shifts is
crucial for a complete and more precise description for
$U_{n/p}(\rho ,\delta ,k)$, and thus for $E_{sym}(\rho )$ and
$L(\rho )$. Experimentally, information on the momentum dependence
of $U_{sym,2}(\rho _{0},k)$ can be in principle obtained from the
isospin dependent nucleon optical model potentials. On the other
hand, all analyses in the present work are based on the
non-relativistic models, it will be thus interesting to see how our
results change in relativistic covariant energy-density functionals,
such as the relativistic mean field models. These studies are in
progress.

\section*{ACKNOWLEDGMENTS}

This work was supported in part by the National Natural Science Foundation
of China under Grant Nos. 10735010, 10775068, 10805026, 10975097, 11135011
and 11175085, Shanghai Rising-Star Program under grant No. 11QH1401100,
``Shu Guang" project supported by Shanghai Municipal Education Commission
and Shanghai Education Development Foundation, the Program for Professor of
Special Appointment (Eastern Scholar) at Shanghai Institutions of Higher
Learning, the National Basic Research Program of China (973 Program) under
Contract No. 2007CB815004, the US National Science Foundation grant
PHY-0757839, the Texas Coordinating Board of Higher Education grant
No.003565-0004-2007, the National Aeronautics and Space Administration under
grant NNX11AC41G issued through the Science Mission Directorate, and the
Research Fund of Doctoral Point (RFDP), No. 20070284016.

\appendix*

\section{Models for nuclear effective interactions}

\label{secmodel}

For completeness, we briefly introduce in this Appendix the nuclear
interaction models used in this work and also present some important
expressions. These models include the isospin- and momentum-dependent MDI
interaction, the Hartree-Fock approach based on Skyrme interactions, and the
Hartree-Fock approach based on the finite-range Gogny interactions. These
models have been extensively used in nuclear structure studies and transport
model simulations for heavy ion collisions.

\subsection{Isospin- and momentum-dependent MDI interaction}

The isospin- and momentum-dependent MDI interaction is a phenomenological
effective interaction based on a modified finite range Gogny interaction
\cite{Das03,Che05}. In the MDI interaction, the potential energy density $%
\varepsilon _{pot}(\rho ,\delta )$ of an asymmetric nuclear matter at total
density $\rho $ and isospin asymmetry $\delta $ is given by%
\begin{widetext}
\begin{equation}
\varepsilon _{pot}(\rho ,\delta )=\frac{A_{u}(x)\rho _{n}\rho _{p}}{\rho _{0}%
}+\frac{A_{l}(x)}{2\rho _{0}}(\rho _{n}^{2}+\rho _{p}^{2})+\frac{B}{\sigma +1%
}\frac{\rho ^{\sigma +1}}{\rho _{0}^{\sigma }}(1-x\delta ^{2})+\frac{1}{\rho
_{0}}\sum_{\tau ,\tau ^{\prime }}C_{\tau ,\tau ^{\prime }}\int \int d\bm pd%
\bm p^{\prime }\frac{f_{\tau }(p)f_{\tau ^{\prime }}(p^{\prime })}{1+(\bm p-%
\bm p^{\prime})^2/{\Lambda ^{2}}},
\end{equation}%
\end{widetext}
where $A_{u}(x)=-95.98-x\frac{2B}{\sigma +1}$ (MeV), $A_{l}(x)=-120.57+x%
\frac{2B}{\sigma +1}$ (MeV), $B=106.35$ (MeV), $\sigma =4/3$, $C_{\tau ,\tau
}=-11.70$ (MeV), $C_{\tau ,-\tau }=-103.40$ (MeV), and $\Lambda =\hbar (3\pi
^{2}\rho _{0}/2)^{1/3}$ are obtained from fitting the momentum dependence of
single-nucleon potential to that predicted by the Gogny Hartree-Fock and/or
the Brueckner-Hartree-Fock calculations, the saturation properties of
symmetric nuclear matter and the symmetry energy of $30.5$ MeV at nuclear
matter saturation density $\rho _{0}=0.16$ fm$^{-3}$. The incompressibility
for cold symmetric nuclear matter at saturation density $\rho _{0}$ is set
to be $K_{0}=211$ MeV. The $x$ parameter in the MDI interaction is
introduced to vary the density dependence of the nuclear symmetry energy
while keeping other properties of the nuclear equation of state fixed \cite%
{Che05}, and it can be adjusted to mimic the predictions of microscopic
and/or phenomenological many-body theories on the density dependence of
nuclear matter symmetry energy. We would like to point out that the MDI
interaction has been extensively used in the transport model for studying
isospin effects in intermediate energy heavy-ion collisions induced by
neutron-rich nuclei \cite%
{LiBA04a,Che04,Che05,LiBA05a,LiBA05b,LiBA06b,Yon06a,Yon06b,Yon07}, the study
of the thermal properties of asymmetric nuclear matter~\cite{Xu07,Xu07b},
and the compact star physics~\cite{XuJ09,XuJ10,XuJ10b}. In particular, the
isospin diffusion data from NSCL/MSU have constrained the value of $x$ to
between $0$ and $-1$ for nuclear matter densities less than about $1.2\rho
_{0}$ \cite{Che05}.

In mean-field approximation, the single-nucleon potential for a nucleon with
momentum $p$ and isospin $\tau $ in asymmetric nuclear matter can be
expressed as \cite{Das03,Che07,Che09}
\begin{widetext}
\begin{eqnarray}
&&U_{\tau }(\rho ,\delta ,p)=A_{u}(x)\frac{\rho _{-\tau }}{\rho _{0}}%
+A_{l}(x)\frac{\rho _{\tau }}{\rho _{0}}+B(\frac{\rho }{\rho _{0}})^{\sigma
}(1-x\delta ^{2})-4\tau x\frac{B}{\sigma +1}\frac{\rho ^{\sigma -1}}{\rho
_{0}^{\sigma }}\delta \rho _{-\tau }  \notag \\
&+&\frac{2C_{\tau ,\tau }}{\rho _{0}}\frac{2}{h^{3}}\pi \Lambda ^{3}\Big[%
\frac{p_{F_{\tau }}^{2}+\Lambda ^{2}-p^{2}}{2p\Lambda }ln\frac{(p+p_{F_{\tau
}})^{2}+\Lambda ^{2}}{(p-p_{F_{\tau }})^{2}+\Lambda ^{2}}+\frac{2p_{F_{\tau
}}}{\Lambda }-2arctan\frac{p+p_{F_{\tau }}}{\Lambda }+2arctan\frac{%
p-p_{F_{\tau }}}{\Lambda }\Big]  \notag \\
&+&\frac{2C_{\tau ,-\tau }}{\rho _{0}}\frac{2}{h^{3}}\pi \Lambda ^{3}\Big[%
\frac{p_{F_{-\tau }}^{2}+\Lambda ^{2}-p^{2}}{2p\Lambda }ln\frac{%
(p+p_{F_{-\tau }})^{2}+\Lambda ^{2}}{(p-p_{F_{-\tau }})^{2}+\Lambda ^{2}}+%
\frac{2p_{F_{-\tau }}}{\Lambda }-2arctan\frac{p+p_{F_{-\tau }}}{\Lambda }%
+2arctan\frac{p-p_{F_{-\tau }}}{\Lambda }\Big],
\end{eqnarray}%
where $\rho _{\tau }=\rho (1+\tau \delta )/2$ and $p_{F_{\tau
}}=\hbar (3\pi ^{2}\rho _{\tau })^{1/3}$.

\subsection{Skyrme Hartree-Fock approach}

For the Skyrme interaction, we use the standard form \cite{Chaba97a} that
has been shown to be very successful in describing the structure of finite
nuclei. Neglecting the spin-orbit interaction term which is irrelevant in
nuclear matter calculations considered here, the nuclear effective
interaction in the standard Skyrme interaction is taken to have a
zero-range, density- and momentum-dependent form \cite{Chaba97a}, i.e.,
\begin{eqnarray}
V_{12}^{Skyrme}(\bm r_{1},\bm r_{2}) &=&t_{0}(1+x_{0}P_{\sigma })\delta (\bm %
r)+\frac{1}{2}t_{1}(1+x_{1}P_{\sigma })[\bm P^{\prime 2}\delta (\bm %
r)+\delta (\bm r)\bm P^{2}]+t_{2}(1+x_{2}P_{\sigma })\bm P^{\prime }\cdot
\delta (\bm r)\bm P  \notag \\
&+&\frac{1}{6}t_{3}(1+x_{3}P_{\sigma })[\rho (\bm R)]^{\alpha }\delta (\bm %
r),
\end{eqnarray}%
where $\bm r=\bm r_{1}-\bm r_{2}$, $\bm R=(\bm r_{1}+\bm r_{2})/2$, $%
P_{\sigma }$ is spin exchange operator, $\bm P=\frac{1}{2i}(\bm\nabla _{1}-%
\bm\nabla _{2})$ is relative momentum operator acting on the right and $\bm %
P^{\prime }$ is its conjugate which acts on the left. The $t_{0}$, $x_{0}$, $%
t_{1}$, $x_{1}$, $t_{2}$, $x_{2}$, $t_{3}$, $x_{3}$ and $\alpha $ are the $9$
Skyrme interaction parameters determined from fitting the binding energies,
charge radii, and other properties of a large number of nuclei in the
periodic table. In the Skyrme Hartree-Fock approach, the single-nucleon
potential in asymmetric nuclear matter is given by \cite{Chaba97a}
\begin{eqnarray}
U_{\tau }(\rho ,\delta ,k) &=&\frac{k^{2}}{8}\rho \lbrack
t_{1}(2+x_{1})+t_{2}(2+x_{2})]+\frac{k^{2}}{8}\rho _{\tau
}[t_{2}(1+2x_{2})-t_{1}(1+2x_{1})]+\frac{1}{2}t_{0}[(2+x_{0})\rho
-(2x_{0}+1)\rho _{\tau }]  \notag \\
&+&\frac{1}{12}t_{3}\rho ^{\alpha }[(2+x_{3})\rho -(2x_{3}+1)\rho _{\tau }]+%
\frac{\alpha }{24}t_{3}\rho ^{\alpha -1}[(2+x_{3})\rho ^{2}-(2x_{3}+1)(\rho
_{n}^{2}+\rho _{p}^{2})]  \notag \\
&+&\frac{1}{8}[t_{1}(2+x_{1})+t_{2}(2+x_{2})](\frac{p_{F_{n}}^{5}}{5\pi
^{2}\hbar ^{5}}+\frac{p_{F_{p}}^{5}}{5\pi ^{2}\hbar ^{5}})+\frac{1}{8}%
[t_{2}(2x_{2}+1)-t_{1}(2x_{1}+1)]\frac{p_{F_{\tau }}^{5}}{5\pi ^{2}\hbar ^{5}%
}.
\end{eqnarray}

\subsection{Gogny interaction}

Gogny interaction has been proved to be very successful in describing not
only nuclear structure but also nuclear matter \cite{Gog77,Dec80}. Compared
with the Skyrme interaction which only contains $\delta $ forces, Gogny
interaction is featured by two finite range terms plus one $\delta $ force
that can well mimic the nucleon-nucleon effective interaction. Neglecting
the spin-orbit interaction term, the conventional Gogny interaction is given
by \cite{Gog77,Dec80}
\begin{equation}
V_{12}^{Gogny}(\bm r_{1},\bm r_{2})=\sum_{i=1}^{2}(W_{i}+B_{i}P_{\sigma
}-H_{i}P_{\tau }-M_{i}P_{\sigma }P_{\tau })e^{-\bm r^{2}/\mu
_{i}^{2}}+t_{0}(1+x_{0}P_{\sigma })[\rho (\bm R)]^{\alpha }\delta (\bm r),
\end{equation}%
where $W_{1}$, $B_{1}$, $H_{1}$, $M_{1}$, $\mu _{1}$, $W_{2}$, $B_{2}$, $%
H_{2}$, $M_{2}$, $\mu _{2}$, $t_{0}$, $x_{0}$ and $\alpha $ are the
$13$ Gogny interaction parameters, and $P_{\tau }$ is the isospin
exchange operator. By using Hartree-Fock approach, we explicitly
write down its single-nucleon potential
\begin{eqnarray}
U_{\tau }(\rho ,\delta ,k) &=&\rho \sum_{i=1}^{2}\pi ^{\frac{3}{2}}\mu
_{i}^{3}(W_{i}+\frac{B_{i}}{2})-\rho _{\tau }\sum_{i=1}^{2}\pi ^{\frac{3}{2}%
}\mu _{i}^{3}(H_{i}+\frac{M_{i}}{2})+t_{0}\rho ^{\alpha }\Big[(1+\frac{x_{0}%
}{2})\rho -(\frac{1}{2}+x_{0})\rho _{\tau}\Big]  \notag  \label{U_tau Gogny} \\
&+&\frac{1}{8}t_{0}\alpha \rho ^{\alpha +1}[3-(2x_{0}+1)\delta
^{2}]+\sum_{i=1}^{2}Z_{i}(k,\tau )(-W_{i}-2B_{i}+H_{i}+2M_{i})  \notag \\
&+&\sum_{i=1}^{2}Z_{i}(k,-\tau)(H_{i}+2M_{i}),
\end{eqnarray}%
with
\begin{equation}
Z_{i}(k,\tau )=\frac{1}{\sqrt{\pi }\mu _{i}k}\Big[e^{-\mu
_{i}^{2}(k+k_{F_{\tau }})^{2}/4}-e^{-\mu _{i}^{2}(k-k_{F_{\tau }})^{2}/4}%
\Big]+\frac{1}{2}\Big\{erf[\mu _{i}(k+k_{F_{\tau }})/2]-erf[\mu
_{i}(k-k_{F_{\tau }})/2]\Big\},
\end{equation}%
where $erf(x)$ is error function.
\end{widetext}

\end{document}